\documentclass[lettersize,journal]{IEEEtran}
\pdfoutput=1
\usepackage{booktabs}
\usepackage{graphicx}
\usepackage{subfigure}
\usepackage{amsmath}
\usepackage{amsthm,amsmath,amssymb}
\usepackage{xcolor}         % colors
\usepackage{color}
\usepackage{amssymb}
\usepackage{bbding}
\usepackage{makecell}
\usepackage{wrapfig}
\usepackage{amsmath}
\usepackage{mathrsfs}
\usepackage[ruled,norelsize]{algorithm2e}
\usepackage{pifont}
\usepackage{cite}
\usepackage{circledsteps}
\usepackage{float}
\usepackage{amsthm,amsmath,amssymb}

\usepackage{mathrsfs}
\ifCLASSINFOpdf

\else

\fi

\usepackage[ruled]{algorithm2e}
\usepackage{booktabs} 
\usepackage{stfloats}
\usepackage{lipsum}

\usepackage{url}

\begin{document}

\title{Dynamic Routing for Integrated
Satellite-Terrestrial Networks: A Constrained Multi-Agent Reinforcement Learning Approach
}

%\iffalse
\author{
        Yifeng~Lyu,
        Han~Hu,~\IEEEmembership{Member,~IEEE,} 
        Rongfei~Fan,~\IEEEmembership{Member,~IEEE,}
        Zhi~Liu,~\IEEEmembership{Senior Member,~IEEE,}
        Jianping~An,~\IEEEmembership{Member,~IEEE,}% <-this % stops a space
        ~and~Shiwen~Mao,~\IEEEmembership{Fellow,~IEEE}
        
\thanks{}        
     
}
%\fi  

\IEEEtitleabstractindextext{%
\begin{abstract}
The abstract goes here.
\end{abstract}

}

% make the title area
\maketitle

\begin{abstract}
The integrated satellite-terrestrial network (ISTN) system has experienced significant growth, offering seamless communication services in remote areas with limited terrestrial infrastructure. However, designing a routing scheme for ISTN is exceedingly difficult, primarily due to the heightened complexity resulting from the inclusion of additional ground stations, along with the requirement to satisfy various constraints related to satellite service quality. To address these challenges, we study packet routing with ground stations and satellites working jointly to transmit packets, while prioritizing fast communication and meeting energy efficiency and packet loss requirements. Specifically, we formulate the problem of packet routing with constraints as a max-min problem using the Lagrange method. Then we propose a novel constrained Multi-Agent reinforcement learning (MARL) dynamic routing algorithm named CMADR, which efficiently balances objective improvement and constraint satisfaction during the updating of  policy and Lagrange multipliers. Finally, we conduct extensive experiments and an ablation study using the OneWeb and Telesat mega-constellations. Results demonstrate that CMADR  reduces the packet delay by a minimum of  $21\%$ and $15\%$, while  meeting stringent energy consumption and packet loss rate constraints, outperforming several baseline algorithms.
\end{abstract}


\IEEEpeerreviewmaketitle

\section{Introduction}

The popularity of LEO satellites is driven by their capability to provide wide-area coverage and high-speed internet communication services, delivering fast and reliable data transmission and communication capabilities to users, particularly in remote areas with limited infrastructure. Telesat~\cite{Telesat} is a satellite communication project with a long-term strategy of deploying 351 satellites operating in the Ka-band at an altitude of approximately 1015 kilometers. Another typical example, OneWeb~\cite{OneWeb} plans to deploy around 720 satellites, enabling a vast satellite network that offers users worldwide reliable and low-latency internet connectivity, delivering high-speed broadband internet services. 

Integrated satellite-ground routing is highly significant as it  ensures reliable communication services that meet the diverse requirements of various application scenarios, such as remote communication, internet access, and emergency response. However, designing a reasonable routing scheme for each satellite and ground station is quite difficult as there are at least two aspects that need to be taken into account. Firstly, the inclusion of ground stations adds an additional dimension to the overall routing strategy space, thereby making the problem more challenging. Secondly, the inherent contradiction between optimized objectives and various constraints in real communication environments makes it difficult to achieve a balance between optimization and constraint satisfaction. 

For the first challenge, most existing  algorithms  overlook the equally critical routing of the uplink and downlink segments involving ground stations~\cite{7572177}~\cite{Jia:Topology}. Few are the algorithms~\cite{9796886} that consider ground-to-space routing, which primarily rely on the distance of sat-to-ground links and available connection time for decision-making, yet fail to fully account for traffic and network environment at the current moment. For the second challenge, existing algorithms~\cite{HUANG2023284}~\cite{9266059} fail to explicitly incorporate constraints into the optimization problem. Instead, they just evaluate the current routing policy based on the extent to which objectives are achieved and constraints are violated, without providing theoretical guarantees that any enhancements to the existing strategy can effectively optimize the objective function while minimizing constraint violations.

In this paper, we study the dynamic routing scheme and address the technical issues mentioned
above. We concentrate  on satellites and ground stations working jointly in the ISTN system  to provide fast communication services while also ensuring compliance with constraints related to energy consumption and packet loss rate. By designing routing tables with structures similar to those of satellites, we seamlessly incorporated ground stations into the ISTN network, enabling distributed decision-making and effectively resolving the first challenge. To address the second challenge, we establish separate critic network structures for different constraints, allowing for dynamic adjustment of the routing strategy to optimize packet delay while ensuring compliance with the specified constraints.

In particular, we formulate the problem of
packet routing with constraints as a  max-min problem and design a CMADR routing algorithm to  iteratively update the routing strategy and Lagrange multiplier of each satellite or ground station based on constraint violations. Extensive simulations are conducted and the results show that CMADR can reduce at least 21\% and 15\% of the average packet delay while satisfying relevant constraints. Our main contributions in this paper are the following:
\begin{itemize}
    \item We consider an  ISTN system that encompasses not only satellites but also crucial ground stations, jointly routing packets in a distributed manner to minimize average packet delay while adhering to energy-efficient  and packet loss rate constraints. 
    \item We formulate it as a  max-min problem and design a constrained MARL algorithm named CMADR to solve it by  updating the distributed routing strategies and Lagrange multipliers to balance between reducing latency and adhering to constraints.
    \item We conduct comprehensive simulations, comparing CMADR with baseline schemes and conducting an ablation study, which unequivocally demonstrated its superiority in performance.
\end{itemize}

The remainder of this paper is organized as follows: In Section II, we introduce  the related work. In Section III, we present the system model for ISTN and formulate the dynamic routing  as an optimization problem. In Section IV, we transform the optimization into a Dec-POMDP problem and propose the CMADR algorithm  to solve it. In Section V, we conduct extensive experiments utilizing various settings to show the efficacy of our proposed algorithm. In Section VI, we conclude this paper.

\section{RELATED WORK}
Routing algorithms can be categorized into two main types: static routing algorithms and dynamic routing algorithms.
\subsection{Static Routing Algorithms}

The core concept involves employing a network topology strategy based on predictable satellite orbit patterns to mitigate the effects of satellite mobility. Offline routing calculations are then performed using static topological algorithms to generate routing tables. Common static routing algorithms include virtual topology-based and virtual node-based approaches.

\textbf{Virtual node}  divide the Earth's surface into zones, assigns virtual satellites to each zone, and maps actual satellites to their corresponding virtual counterparts. During satellite handover, routing information is passed from the previous satellite to the next, converting the routing problem into finding optimal routes in a static network~\cite{Mauger}~\cite{Ekici}. %Chen \textit{et al.} implements an enhanced VN method that utilizes Celestial Sphere Division (CSD-VN) to mitigate the effects of Earth rotation, incorporating a virtual node mapping update model to handle asynchronous switching caused by phase differences among neighboring orbital satellites, in a single-layer mega-constellation~\cite{8926396}.

\textbf{Virtual topology} algorithms leverage the periodicity of satellite operations to divide the system cycle of LEO constellation networks into time slots. Each slot represents a static and unchanging satellite network topology, transforming the dynamic topology into a series of repeating static structures. Routing tables for these static structures are calculated using algorithms like Dijkstra's~\cite{553679} and stored on the satellites~\cite{lu2013virtual}~\cite{Jia:Topology}~\cite{WANG200743}. During routing computations, satellites select the corresponding virtual topology based on the current time and utilize pre-calculated routing tables for lookups~\cite{7275422}~\cite{NSD}~\cite{9796886}. %For example, the Finite State Automaton (FSA)~\cite{chang1998fsa} routing algorithm treats the changing LEO satellite network topology as a state transition process, optimizing link resource allocation within a finite state machine based on virtual topology. 

However, considering the real-time traffic and link status changes in satellite networks, static routing strategies are unable to effectively cope with complex network environments.

\subsection{Dynamic Routing Algorithms}

To address the issue of inflexible static routing, recent dynamic routing algorithms, particularly those leveraging Artificial Intelligence and Machine Learning (AI/ML) in the similar domain, dynamically recalculate routes based on the topology and collected link state information to adapt to changes in the network~\cite{cao2022dynamic}.

Huang \textit{et al.}~\cite{HUANG2023284} introduced QRLSN, a Q-learning-based~\cite{Q-learning} dynamic distributed routing scheme that employs multi-objective optimization to discover an efficient routing strategy, minimizing both end-to-end delay and network traffic overhead load. Liu \textit{et al.}~\cite{9266059} introduced DRL-ER, a deep Q network (DQN)-based~\cite{DQN} energy-efficient routing protocol. This protocol aims to balance satellite battery energy and ensure bounded end-to-end delay while enabling satellites to learn a well-balanced energy usage routing policy. Xu et al.~\cite{9882125} presented a fully distributed routing algorithm incorporating spatial location information based on Multi-Agent deep Reinforcement Learning (FDR-MARL) for large-scale satellite networks, aiming to minimize average delivery time. However, these schemes do not explicitly address the optimization of routing involving heterogeneous ground stations, focusing solely on satellites. Zhang \textit{et al.}~\cite{10012942} integrated the mean field theory~\cite{9137257} to illustrate the interaction among agents and their neighbors and subsequently formulated the conventional DQN for training individual satellites and ground stations. However, without establishing clear relationships between the different constraints and routing strategies in optimizing the targeted metric, there is no assurance of performance. In this paper, we aim to tackle these issues by introducing a dynamic routing scheme named CMADR, distinguishing it from existing static and dynamic schemes.

\section{SYSTEM MODEL AND PROBLEM FORMULATION}

We start by presenting an overview of the architecture for the ISTN. Following that, we define the networking model and elaborate on the various components such as communication delay, energy consumption, and packet loss rate with detailed explanations. Finally, we formulate the problem  and summarize notations in Table I.
\begin{table}[htbp]
	\centering
	\caption{NOTATIONS AND CORRESPONDING DESCRIPTIONS.}
	\begin{tabular}{|c|c|}
		\hline
		\textbf{Notation}&\textbf{Description} \\ 
		\hline  

		${\mathcal{G}}$&Graph of the ISTN \\
		${G}$&Set of ground stations\\
		${S}$&Set of satellites \\  
		${P}$&Set of packets \\  

		${T}$&Total time horizon  \\
		${W}$&Total time slots of ${T}$   \\	
        ${{V}_{i}^{t}}$ &Neighbor satellites and connectable ground stations of  ${S}_{i}$\\	
        ${{V}_{j}^{t}}$ & Connectable satellites of ${G}_{j}$ \\	  
        
         ${D}_{i,{i}^{'}}^{t}$, ${C}_{i,{i}^{'}}^{t}$&Distance and transmit rate from ${S}_{i}$ to ${S}_{i^{'}}$ \\
        
         ${D}_{j,{i^{'}}}^{t}$, ${C}_{j,{i^{'}}}^{t}$&Distance and transmit rate from ${G}_{j}$ to ${S}_{i^{'}}$ \\

        ${P}_{R}, {P}_{O}$&Transmission power for the RF link and  the FSO link  \\

        ${B}_{R}, {B}_{O}$&Bandwidth for the RF link and  the FSO link  \\

        ${E}_{i}^{t}$, ${E}_{j}^{t}$&Energy consumption of ${S}_{i}$ and ${G}_{j}$\\

        ${E}_{G}, {E}_{S}$&Energy limit of each ground station/satellite \\

        ${D}_{m}$ & End-to-end packet delay of $P_m$ 	 \\
        
        ${D}_{m}^{T}$, ${D}_{m}^{P}$& Transmission delay and propagation delay of $P_m$ 	 \\

        ${D}_{m}^{Q}$, ${D}_{m}^{C}$& Queuing delay and processing delay of $P_m$ 	 \\

        ${L}_{P}$& Packet size \\
        
        $H$&Signal propagation velocity\\

        $P_L$ &Packet loss rate threshold\\

		\hline 

	\end{tabular}

\end{table}
\vspace{-10pt}

\subsection{Overall Architecture}

\begin{figure*}
\centering  
\includegraphics[height=5.5cm,width=17cm]{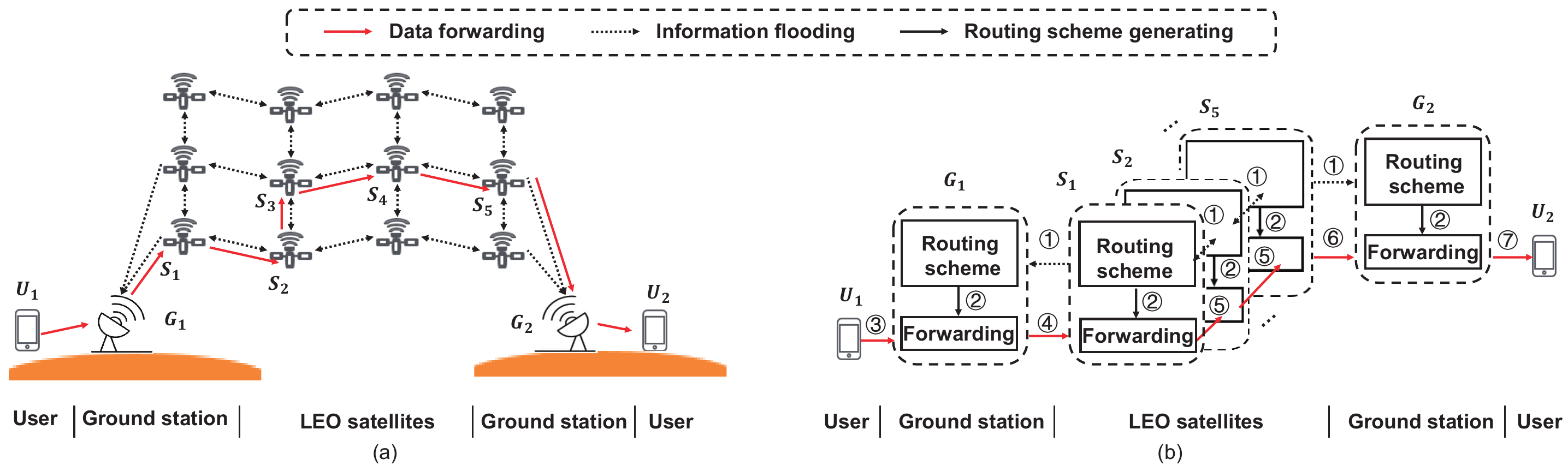}
\vspace{-10pt}
\caption{(a) The architecture of ISTN; (b) Workflow of the architecture: \ding{172} During each time slot, every satellite distributes its own state information to four neighboring satellites and reachable ground stations; \ding{173} Every satellite or ground station generates a routing scheme based on its respective states information, as well as information received from external sources; \ding{174} Within each region serviced by a ground station, user messages are packetized and stored in the station's buffer; \ding{175} Based on the routing scheme, the ground station with data packets in the buffer uploads them to the satellite;  \ding{176} In accordance with their respective routing schemes, every satellite transmits data to the next satellite in line;  \ding{177} Data is transmitted to the ground station;  \ding{178} The delivery of data is directed towards the designated user.} 
\vspace{-10pt}
\end{figure*}

In this study, both ground stations and LEO satellites are considered to be indispensable conduits that provide seamless connectivity to users, as depicted in Fig. 1(a). Each ground station receives data from users and transmits it to satellites. The satellites then relay the data to other satellites before transmitting it back down to ground stations. Finally, ground stations forward the data to the designated destination users.

As illustrated in Fig. 1(b), the ISTN operates through a detailed communication process, which can be broken down into the following steps: Firstly, each satellite periodically transmits its own information to four neighboring satellites, as well as any nearby ground stations for information sharing purposes. Then, both  ground stations and satellites formulate routing schemes to guide routing procedures. Each ground station receives user messages, packetizes them, and enqueues the messages into the buffer of the ground station. Each packet is then transmitted to the next satellite in a first-in-first-out order, based on the destination and routing scheme. If a ground station's service area covers the intended destination of the packet, it is transmitted directly to the ground station. Lastly, the ground station forwards the received packet to the designated recipient.

Our goal is to design an efficient distributed dynamic network routing strategy for both satellites and ground stations to provide high-quality communication services. However, achieving this goal poses significant challenges due to the following factors. Firstly, the decision-making process in the ISTN adds complexity to obtaining a jointly optimal routing scheme. The coordination between satellites and ground stations requires careful consideration to ensure efficient routing decisions. Additionally, the limited energy consumption capabilities of both ground stations and satellites present a challenge. The routing scheme must take into account the energy constraints and optimize the usage to prolong the operational lifetime of these nodes. Similarly, the restricted buffer capacity of both ground stations and satellites necessitates careful management to avoid congestion and ensure smooth communication.

\subsection{Networking Model}

We represent the set of ground stations as  $G = \{{G}_{j}|j = 1, ..., {N}_{G}\}$ where ${N}_{G}$ denotes the total number of ground stations and ${G}_{j}$ signifies the j-th station. Similarly, we denote the set of satellites as $S = \{{S}_{i}|i = 1, ..., {N}_{S}\}$ where ${N}_{S}$ represents the total number of satellites and ${S}_{i}$ indicates i-th satellite. The inter-satellite connections adhere to the ubiquitous +Grid topology~\cite{9351765}~\cite{1281968}, wherein each satellite maintains fixed links with its preceding and succeeding neighbors within the same orbital plane, along with the two corresponding satellites situated in adjacent orbital planes. Then inter-satellite links remain constant all the time while  between satellite-terrestrial links change all the time. 

For convenience, we divide the service time  $T$ into a total of $W$ time slots, following the principle of virtual topology ~\cite{9222519}~\cite{9944375}. Within each time slot,  links between satellites and ground stations also remain constant and will change at the next time slot. To accurately represent the relationship of the network, we define ${{V}_{i}^{t}}$ as the set of four neighbor satellites and connectable ground stations of  ${S}_{i}$  at time slot $t$. Similarly, we define ${{V}_{j}^{t}}$ as the set of connectable satellites of ${G}_{j}$  at time slot $t$. Then the set of all ${{V}_{i}^{t}}$ and ${{V}_{j}^{t}}$ is denoted as $V$ and  the ISTN system can be represented mathematically as an undirected graph $\mathcal{G} = 〈G, S, V〉$. To depict the routing scheme that involves packet forwarding by distributed stations and satellites, hop-by-hop, we have categorized the links in the integrated sat-ground network into three types: \textbf{uplinks} from ground stations to satellites, \textbf{inter-satellite links} from one satellite to another, and \textbf{downlinks} from satellites to ground stations. Packets forwarded from ground station ${G}_{j}$ to satellite  ${S}_{{i}^{'}}$ in time slot $t$ are denoted as ${P}_{j, {i}^{'}}^{t}$ for uplinks, where ${S}_{{i}^{'}} \in {V}_{j}^{t}$. Similarly, for inter-satellite links,  packets forwarded from satellite ${S}_{i}$ to satellite ${S}_{{i}^{'}}$ are denoted as  ${P}_{i, {i}^{'}}^{t}$, where ${S}_{{i}^{'}} \in {V}_{i}^{t}$. For downlinks,  packets forwarded from satellite  ${S}_{i}$ to  ground station ${G}_{{j}^{'}}$ are  denoted as ${P}_{i, {j}^{'}}^{t}$, where ${S}_{{j}^{'}} \in {V}_{i}^{t}$. We define the set of all  packets as $\mathcal{P} = \{{P}_m|m=1, …,{N}_{P}\}$ where ${N}_{P}$ represents the total number of packets and ${P}_{m}$ denotes the m-th packet.

\subsection{Communication Delay}
The quality of service in a communication network system is mainly measured by the communication delay. The end-to-end delay ${D}_{m}$  for a particular packet ${P}_{m}$ can be calculated using the equation given in~\cite{6222005}, which is as follows:
\begin{equation}
{D}_{m} = 
{D}_{m}^{Q} + 
{D}_{m}^{C} + 
{D}_{m}^{T} + 
{D}_{m}^{P}.
\end{equation} 
Here, ${D}_{m}^{Q}$ represents  the queuing delay, ${D}_{m}^{C}$ denotes the processing delay, ${D}_{m}^{T}$ represents the transmission delay,  and  ${D}_{m}^{P}$ denotes the propagation delay.  It is important to note that the delay of packets from users to the ground station and vice versa are not included in the ISTN system, and hence, they are not taken into account.

\textbf{Queuing delay}. When a packet arrives at the buffer of either a satellite or a ground station, it is required to wait until all the preceding packets have been transmitted. The queuing delay ${D}_{m}^{Q}$ of the specific packet ${P}_{m}$ is influenced by the communication demand from users and the routing scheme of all satellites and ground stations in the network.

\textbf{Processing delay}. Each packet must be unpacked to obtain its destination address and lookup the routing table to determine the next transfer location. The processing delay for each ground station or satellite is assigned to a constant time denoted as ${D}_{L}$~\cite{6222005}. Therefore, the processing delay ${D}_{m}^{C}$ for packet ${P}_{m}$ is determined by multiplying a constant processing time ${D}_{L}$ for each satellite or ground station with the number of forwarding hops in the transmission path:
\begin{align}
&{D}_{m}^{C} = \sum_{t=1}^{W}\Bigg(\sum_{j=1}^{{N}_{G}}\sum_{{S}_{{i}^{'}} \in {V}_{j}^{t}}\sum\limits_{{P}_{m} \in {P}_{j, {i}^{'}}^{t}}{D}_{L}  +  \nonumber \\  
& \quad \sum_{i=1}^{{N}_{S}}\sum_{{S}_{{i}^{'}} \in {V}_{i}^{t}}\sum\limits_{{P}_{m} \in {P}_{i, {i}^{'}}^{t}} {D}_{L}  + \sum_{i=1}^{{N}_{S}}\sum_{{G}_{{j}^{'}} \in {V}_{i}^{t}}\sum\limits_{{P}_{m} \in {P}_{i, {j}^{'}}^{t}}  {D}_{L} \Bigg)
\end{align} 
Here, the processing delay of the uplink is represented by $\sum_{j=1}^{{N}_{G}}\sum_{{S}_{{i}^{'}} \in {V}_{j}^{t}}\sum\limits_{{P}_{m} \in {P}_{j, {i}^{'}}^{t}}{D}_{L}$,  while  that of inter-satellite links is indicated by $\sum_{i=1}^{{N}_{S}}\sum_{{S}_{{i}^{'}} \in {V}_{i}^{t}}\sum\limits_{{P}_{m} \in {P}_{i, {i}^{'}}^{t}} {D}_{L}$. Similarly,  $\sum_{i=1}^{{N}_{S}}\sum_{{G}_{{j}^{'}} \in {V}_{i}^{t}}\sum\limits_{{P}_{m} \in {P}_{i, {j}^{'}}^{t}}  {D}_{L}$ represents the processing delay of the downlink. The sum of  three components, integrated over all time slots, forms the overall processing delay for ${P}_{m}$.

\textbf{Transmission delay}.  Satellites and ground stations transmit packets by converting them from electrical signals to electromagnetic waves or laser signals. In accordance with current practices and the design of the Starlink constellation~\cite{1386877}~\cite{532260}, communication between satellites and ground stations is accomplished using radio-frequency (RF) links, while communication among satellites relies on free-space optical (FSO) links. Taking into account the characteristics of both kinds of links, we assume that all links are free from Doppler effect~\cite{7888557}. Regarding the microwave links connecting satellites and ground stations, we exemplify with the ground station ${G}_{j}$ and the connectable satellite ${S}_{{i}^{'}}$, where ${S}_{{i}^{'}} \in {V}_{j}^{t}$.  The transmission rate ${C}_{j,{i}^{'}}^{t}$ can be  calculated as~\cite{9222519}:
\begin{equation}
{C}_{j,{i}^{'}}^{t} = {B}_{R}\times{log}_{2}\left(1 + \frac{P_R \times {{h}_{j,{i}^{'}}^{t}}^2}{{\sigma}^2}\right),
\end{equation} 
where ${B}_{R}$ denotes  the  frequency  of  links  between ground stations and satellites. ${P}_{R}$ indicates the antenna transmission power for RF links and ${\sigma}^2$ is the power of the background noise. ${h}_{j,{i}^{'}}^{t}$ is the channel gain~\cite{Qu:hard} and can be calculated as 
${h}_{j,{i}^{'}}^{t} = {G}_{T} + {G}_{R} - {L}_{j,{i}^{'}}^{t},$ where ${G}_{T}$ represents the transmitting antenna gain, and ${G}_{R}$ denotes the receiving antenna gain. ${L}_{j,{i}^{'}}^{t}$ represents the signal attenuation during transmission, including free space path loss, atmospheric loss, polarization loss, and antenna misalignment loss. Atmospheric loss, polarization loss, and antenna misalignment loss typically cause less than a 1 dB reduction and can be ignored~\cite{9383778}~\cite{9079470}~\cite{8951285}. The free space loss is calculated as $\left(\frac{4 \pi {f}_{c} {D}_{j, {i}^{'}}^{t}}{H}\right)^2$, where
${D}_{j, {i}^{'}}^{t}$ indicates the distance from ${G}_{j}$ to ${S}_{{i}^{'}}$, and ${f}_{c}$ denotes the central carrier frequency  for Ka-Band, and ${H}$ is the light speed.  Then the transmission delay for a packet from ${G}_{j}$ to ${S}_{{i}^{'}}$   is $\frac{{L}_{P}}{{C}_{j,{i}^{'}}^{t}}$, where ${L}_{P}$ is the length of each packet. The calculation for the downlink process from ${S}_{i}$ to ${G}_{{j}^{'}}$ remains the same. 

For laser links between satellites, we exemplify with the satellite ${S}_{i}$ and neighboring  satellites ${S}_{{i}^{'}}$, where ${S}_{{i}^{'}} \in {V}_{i}^{t}$. Then the transmission rate is denoted as:
\begin{equation}
{C}_{i,{i}^{'}}^{t} = \frac{1}{2}{{B}_{O}}\times{log}_{2}\left(1 + {k}_{1}\cdot{{e}^{-{k}_{2} \cdot {D}_{i,{i}^{'}}^{t}  }}\right),
\end{equation}  
where $k_{1}= 
\frac{\gamma_{{F}}^{2}}{2 \pi e \alpha^{2}}$,  and $k_{2}=2 \beta$. Here, ${B}_{O}$ is bandwidth for FSO links and ${D}_{i,{i}^{'}}^{t}$ is the distance from ${S}_{i}$ to ${S}_{{i}^{'}}$.  $\gamma_{\mathrm{O}}^{2}$ is the average optical SNR (ASNR) and could be calculated as  $\gamma_{\mathrm{O}}^{2}=\frac{{P}_{O}^{2}}{\sigma_{\mathrm{O}}^{2}}$, where ${P}_{O}$ and $\sigma_{\mathrm{O}}^{2}$ are the average optical power and noise variance, respectively~\cite{9944375}. The parameter $\beta$ can be calculated as $\beta=\frac{\beta_{\mathrm{dB}}}{10^{4} \log _{10} e}$, where $\beta_{\mathrm{dB}}=\frac{3.91}{L}\left(\frac{\lambda}{550 }\right)^{-p}$ ~\cite{233}~\cite{5238736} depends on the wavelength $\lambda$ and $L$ denotes the visibility~\cite{233}. The transmission delay for a packet  from ${S}_{i}$ to ${S}_{{i}^{'}}$ is  $\frac{{L}_{P}}{{C}_{i,{i}^{'}}^{t}}$ and the transmission delay ${D}_{m}^{T}$ is calculated as: \begin{align}
&{D}_{m}^{T} = \sum_{t=1}^{W}\Bigg(\sum_{j=1}^{{N}_{G}}\sum_{{S}_{{i}^{'}} \in {V}_{j}^{t}}\sum\limits_{{P}_{m} \in {P}_{j, {i}^{'}}^{t}}\frac{{L}_{P}}{{C}_{j, {i}^{'}}^{t}}  +  \nonumber \\  
& \sum_{i=1}^{{N}_{S}}\sum_{{S}_{{i}^{'}} \in {V}_{i}^{t}}\sum\limits_{{P}_{m} \in {P}_{i, {i}^{'}}^{t}} \frac{{L}_{P}}{{C}_{i, {i}^{'}}^{t}}  + \sum_{i=1}^{{N}_{S}}\sum_{{G}_{{j}^{'}} \in {V}_{i}^{t}}\sum\limits_{{P}_{m} \in {P}_{i, {j}^{'}}^{t}}  \frac{{L}_{P}}{{C}_{i, {j}^{'}}^{t}} \Bigg)
\end{align} 
Here, the transmission delay is also composed of delay for three types of links as the processing delay.

\textbf{Propagation delay}. It is worth noting that irrespective of whether microwave or laser links are employed between satellites and ground stations, the propagation delay remains constant at the speed of light. The propagation delay for the packet ${P}_{m}$ is indicated by ${D}_{m}^{P}$, and can be computed as the cumulative distance along the routing path divided by the propagation velocity $H$~\cite{6222005}:
\begin{align}
&{D}_{m}^{P} = \sum_{t=1}^{W}\Bigg(\sum_{j=1}^{{N}_{G}}\sum_{{S}_{{i}^{'}} \in {V}_{j}^{t}}\sum\limits_{{P}_{m} \in {P}_{j, {i}^{'}}^{t}}\frac{{D}_{j, {i}^{'}}^{t}}{H}  +  \nonumber \\  
& \sum_{i=1}^{{N}_{S}}\sum_{{S}_{{i}^{'}} \in {V}_{i}^{t}}\sum\limits_{{P}_{m} \in {P}_{i, {i}^{'}}^{t}} \frac{{D}_{i, {i}^{'}}^{t}}{H}  + \sum_{i=1}^{{N}_{S}}\sum_{{G}_{{j}^{'}} \in {V}_{i}^{t}}\sum\limits_{{P}_{m} \in {P}_{i, {j}^{'}}^{t}}  \frac{{D}_{i, {j}^{'}}^{t}}{H} \Bigg),
\end{align} 
which can be divided into three types of links as well as transmission delay.

\subsection{Energy Consumption}
The energy consumption during communication services by both ground stations and satellites is mainly attributed to packet transfer, which must be restricted to the predetermined upper limit. As the information exchange between satellites and ground stations occurs infrequently, our study does not consider any resulting energy costs from such periodic flooding. Additionally, as the energy from ground stations to users is not the focus of this paper, it is also disregarded.

\textbf{Energy consumption of ground stations}. The energy consumption of ground stations is attributed  to the uploading of data to satellites. For ${G}_{j}$, the transmission energy for a packet from it to the satellite ${S}_{{i}^{'}}$ at time slot $t$ can be denoted as $\frac{
 {L}_{P}}{{C}_{j, {i}^{'}}^{t}} \times {P}_{R}$, where ${P}_{R}$ is the antenna transmission power for RF links. The accumulated energy consumption of transmitting packets from ${G}_{j}$ should be less than the energy limit  ${E}_{G}$ for each ground station. 

\textbf{Energy consumption of satellites}. The energy consumption of satellites is due to the forwarding packets to neighboring satellites and downloading packets to ground stations. Typically, a satellite's power system utilizes solar panels to generate electricity from solar irradiance, and employs battery cells to store the energy~\cite{7572177}. Constructing and launching a satellite is an expensive endeavor, hence we expect each satellite to operate for as long as possible. To ensure this, the energy consumption of satellites should be supplemented by solar energy as much as possible and minimizing the utilization of the satellite's battery. For the satellite ${S}_{i}$, the forwarding energy  from it to the satellite ${S}_{i}$ at time slot $t$ can be denoted as $\frac{
  {L}_{P}}{{C}_{j, {i}^{'}}^{t}} \times {P}_{O}$ and the downloading energy  from it to the ground station ${G}_{{j}^{'}}$ is $\frac{
 {L}_{P}}{{C}_{j, {i}^{'}}^{t}} \times {P}_{R}$, where ${P}_{O}$ is the antenna transmission power for FSO links. Then the accumulated energy consumption of ${S}_{i}$ is limited to the maximum energy supplement received by the solar panels ${E}_{S}$.

\subsection{Packet Loss Rate}

If the accumulated number of packets exceeds the maximum buffer capacity for each satellite, newly received packets have to be dropped. To ensure high-quality service of communication, our routing scheme must be designed carefully to limit the total packet loss rate ~\cite{126989}~\cite{957312}~\cite{391749}:
\begin{equation}
\quad \qquad   \frac{1}{{N}_{P}}  \sum_{t=1}^{W} \sum_{i=1}^{{N}_{S}} {P}_{i}^{t}  \leq {P}_{L}, 
\end{equation} 
where ${P}_{i}^{t}$ is the number of dropped packets at the satellite ${S}_{i}$ at time slot $t$. The  accumulated value over all time slots and satellites denotes the total number of dropped packets, which must not exceed the predetermined upper limit of packet loss rate $P_L$ to ensure a certain level of communication quality.

\subsection{Problem Formulation}
In order to guarantee prompt and effective communication across the network, it is of utmost importance to minimize the average packet delay. Additionally, taking into account the aforementioned energy and buffer limitations, we have formulated the optimization problem as $P_1$.
\begin{align}
 & \text{(P1)}   \min_{\substack{{P}_{j, {i}^{'}}^{t},  {P}_{i, {i}^{'}}^{t},  {P}_{i, {j}^{'}}^{t} }}  \frac{1}{{N}_{P}}   \sum_{m=1}^{{N}_{P}} \left( {D}_{m}^{Q} + {D}_{m}^{C} + 
{D}_{m}^{T} + 
{D}_{m}^{P} \right)  \\
 & s.t. \quad  \sum_{t=1}^{W}\sum\limits_{m=1}^{{N}_{P}}\sum_{{S}_{{i}^{'}} \in {V}_{j}^{t}} \sum\limits_{{P}_{m} \in {P}_{j, {i}^{'}}^{t}} \frac{
 {L}_{P}}{{C}_{j, {i}^{'}}^{t}} \times {P}_{R}  \leq {E}_{G},   \nonumber \\ 
 &\qquad \qquad \qquad \qquad \qquad \qquad \qquad \qquad j =  1, ..., {N}_{G}, \\
 &\quad   \quad  \sum_{t=1}^{W} \sum\limits_{m=1}^{{N}_{P}}\Bigg(\sum_{{S}_{{i}^{'}} \in {V}_{i}^{t}}\sum\limits_{{P}_{m}  \in {P}_{i, {i}^{'}}^{t}} \frac{ 
 {L}_{P}}{{C}_{i, {i}^{'}}^{t}}\times {P}_{O}  + \nonumber \\ &\qquad \sum_{{G}_{{j}^{'}} \in {V}_{i}^{t}} \sum\limits_{{P}_{m} \in {P}_{i, {j}^{'}}^{t}}\frac{ 
 {L}_{P}}{{C}_{i, {j}^{'}}^{t}} \times {P}_{R}  \Bigg)     \leq {E}_{S},  \nonumber \\ &\qquad \qquad \qquad \qquad \qquad \qquad \qquad \qquad
 i =  1, ..., {N}_{S}, \\
 & \qquad\quad (7) \nonumber. 
 \end{align}
 Regarding the objective function in (8), the average end-to-end delay is determined by the sum of queuing delay, processing delay,  transmission delay and  propagation delay of all packets.  For the constraint (9), $\sum_{t=1}^{W}\sum\limits_{m=1}^{{N}_{P}}\sum_{{S}_{{i}^{'}} \in {V}_{j}^{t}} \frac{
\sum\limits_{{P}_{m} \in {P}_{j, {i}^{'}}^{t}} {L}_{P}}{{C}_{j, {i}^{'}}^{t}} \times {P}_{R}$ denotes the total energy consumed by ${G}_{j}$ for transmitting packets to connectable satellites ${S}_{{i}^{'}}$. This quantity must be lower than the maximum energy ${E}_{G}$ of each ground station. For constraint (10), $\sum_{t=1}^{W} \sum\limits_{m=1}^{{N}_{P}}\sum_{{S}_{{i}^{'}} \in {V}_{i}^{t}} \frac{ 
\sum\limits_{{P}_{m} \in {P}_{i, {i}^{'}}^{t}} {L}_{P}}{{C}_{i, {i}^{'}}^{t}} \times {P}_{O}$  represents the energy consumed in transmitting packets from ${S}_{i}$ to neighbor satellites, while  $  \sum_{t=1}^{W} \sum\limits_{m=1}^{{N}_{P}}  \sum_{{G}_{{j}^{'}} \in {V}_{i}^{t}} \frac{ \sum\limits_{{P}_{m} \in {P}_{i, {j}^{'}}^{t}} {L}_{P}}{{C}_{i, {j}^{'}}^{t}}    \times {P}_{R}  $ denotes the energy consumed in transmitting packets from ${S}_{i}$ to connectable ground stations. Both of these contribute to the total energy consumption of ${S}_{i}$, which should be less than ${E}_{S}$ of each satellite. For constraint (11), the packet loss
rate cannot exceed the upper limit $P_L$. 

Significantly, the difficulty of explicitly describing  packet loss rate in constraint (7) and  queuing delay ${D}_{m}^{Q}$ in equation (8)  makes it challenging for traditional methods such as convex optimization algorithms to solve this problem. More importantly, given the dynamic and ever-changing nature of network conditions, satellites and ground stations are required to adapt their routing strategies based on the environment at each time slot $t$. Then we  formulate this intricate problem as a constrained decentralized partially observable Markov decision process (Dec-POMDP), leveraging a constrained multi-agent reinforcement learning algorithm to solve it.

\section{Constrained Multi-agent Dynamic Routing Algorithm}

\begin{figure*}[htb]
\centering  
\includegraphics[height=5cm,width=17cm]{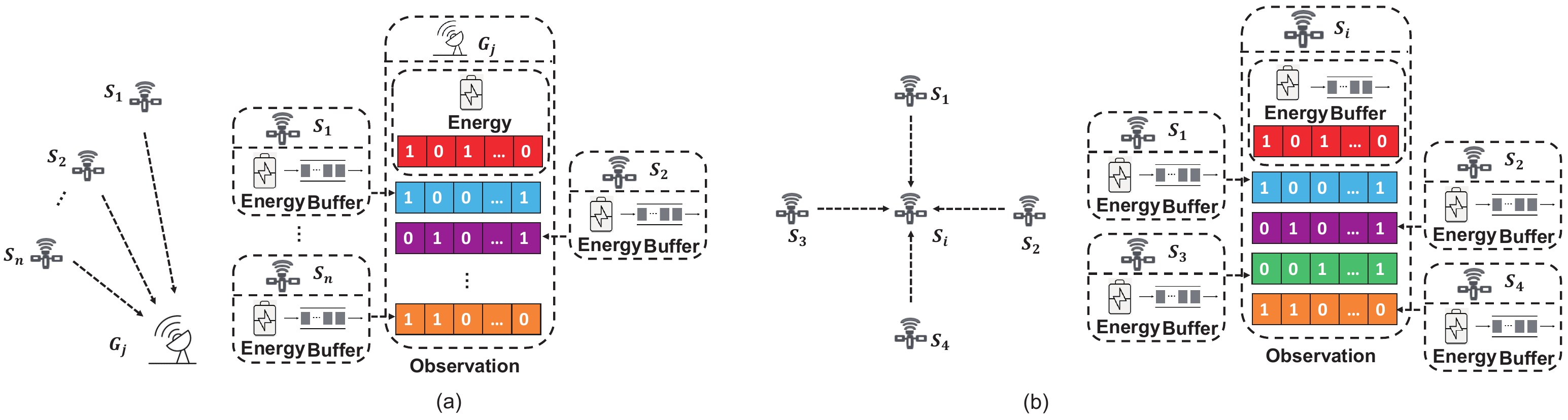}
\vspace{-10pt}
\caption{(a) Each ground station's observation includes  energy consumption and buffer usage of all connectable satellites, as well as the station's own energy consumption; (b)Each LEO satellite's observation is based on its own energy consumption and buffer usage, as well as those of  four neighboring satellites.} 
\vspace{-10pt}
\end{figure*}

We initiate our algorithm by formulating the problem as a constrained Dec-POMDP and then proceed to apply the Lagrangian method, thereby transforming it into a max-min optimization problem. Following this, we introduce the architecture of CMADR. Finally, we offer a comprehensive explanation of the training and execution procedures of CMADR, delving into the intricate details involved.

\subsection{Overview of Dec-POMDP}
\textbf{Dec-POMDP}~\cite{pomdp} is typically defined by a tuple   $\left\langle \mathcal{S}, \mathcal{A},\mathcal{O}, \mathcal{P},  \gamma, \mathcal{R}, \mathcal{C}, \boldsymbol{c}\right\rangle$. Here,  $\mathcal{S}$ is the global state space, and $\mathcal{O}$ is the set of observations that each agent can obtain from the environment. The policy of selecting the next satellite for each station is represented by $\pi_{\theta_j}$, while the policy for each satellite is represented by $\pi_{\theta_i}$. At time slot $t$, each satellite or ground station generates an action ${a}^{t}_{j(i)}$ based on its respective policy $\pi_{\theta_{j(i)}}({a}^{t}_{j(i)}\mid{o}^{t}_{j(i)})$. By taking the joint action  $\vec{a}^t=\prod_{j=1}^{N_G}{a}^{t}_{j}\prod_{i=1}^{N_S} {a}^{t}_{i}$, all agents interact with the environment. They receive the  reward $r^t$ and cost $c^t$ based on this joint action, where $\mathcal{R}$ and $\mathcal{C}$ represent the sets of rewards and costs for all time slots.  The set of joint actions is denoted as $\mathcal{A}$, and $\boldsymbol{c}$ represents limitations that need to be satisfied corresponding to $\mathcal{C}$. Simultaneously, the environment transitions to a new state  ${s}^{t+1} \sim \mathrm{p}\left(\cdot \mid \mathrm{s}_{t}, {a}_{t}\right)$, where  $\mathcal{P}$  is the set of probabilistic transition functions. The discount factor $\gamma \in[0,1)$ is used to discount future rewards. Then we will specifically define the components of state, observation, action, reward, and cost.

\textbf{State}. The global state is formed by concatenating all the observations from ground stations and satellites.

\textbf{Observation}. The specific observation configurations of the satellites and ground stations can be observed in Fig.2. In Fig.2 (a), observation of a ground station consists of two components: its own energy consumption and the environmental information received from connectable satellites, which includes their energy consumption and buffer usage. Similarly, in Fig.2 (b), observation of a satellite comprises two components: its own energy consumption and buffer usage, as well as information received from four neighboring satellites including their energy consumption and buffer usage.

\textbf{Action}. To handle incoming packets, a ground station's task is selecting the next satellite within its service area when presented with a destination address. The size of the ground station's action set is linearly proportional to the number of possible destinations. Similarly, for each satellite, the action set entails choosing the next satellite among four neighboring satellites. The size of the satellite's action set is also linearly proportional to the number of possible destinations.

\textbf{Reward}. The reward in this integrated communication system represents the average time delay experienced by all transmitted packets via satellites and ground stations. As the goal of reinforcement learning is to optimize the expected reward, and the objective function in equation (8) aims to minimize the average time delay of all packets, we ingeniously define the reward as the average transferring rate of all packets. In each time slot, the packet's transmission rate is calculated as the distance it travels towards its destination ground station divided by the corresponding duration, with the distance being measured from the vertical projection of the packet onto the ground to the destination ground station. By improving the accumulated reward through training, the average transmission rate of all packets is increased, resulting in a decrease in the average time delay.

\textbf{Cost.}
There are three constraints that contribute to the overall cost, as outlined in constraints (7), (9), and (10). The packet loss component in constraint (7) is a result of the joint routing scheme. We develop a central cost critic to evaluate the cost of the entire system, with the cost of each step being the total number of dropped packets during the given time slot. On the other hand, the energy consumption costs associated with each ground station and satellite in constraints (9) and (10) arise from the energy consumed by each entity during that particular time slot. We design a local cost critic to evaluate the cost of each individual item and the cost of each step is the accumulated energy consumption of the respective satellite or ground station over the given time slot. 

The goal is to maximize the expected total reward while satisfying the expected total cost and each agent's safety constraints and (P1) can be rewritten as:
\begin{align}
  \quad \text{(P2)} & \quad    J_R({\pi}) \triangleq {E}_{ s \sim p, \vec{a} \sim {\pi} }\left[\sum_{t=1}^{W} \gamma^{t} r\left({s}^{t}, {\vec{a}}^{t}\right)\right]
 \\
 & s.t. \quad  J_C({\pi}) \triangleq {E}_{ s \sim p, \vec{a} \sim {\pi} }\left[\sum_{t=1}^{W} \gamma^{t} c\left({s}^{t}, {\vec{a}}^{t}\right)\right] \leq P_L,    \\ 
 & \qquad J_{j}(\pi) \triangleq {E}_{ {o}_{j} \sim p, {a}_{j} \sim {\pi}_{j} }\left[\sum_{t=1}^{W} \gamma^{t} c\left({o}_{j}^{t}, {a}_{j}^{t}\right)\right] \leq E_G, \quad \nonumber\\ & \qquad\qquad\qquad\qquad\qquad\qquad  j=1,2,...,N_G, \\ 
 & \qquad J_{i}(\pi) \triangleq {E}_{ {o}_{i} \sim p, {a}_{i} \sim {\pi}_{i} }\left[\sum_{t=1}^{W} \gamma^{t} c\left({o}_{i}^{t}, {a}_{i}^{t}\right)\right] \leq E_S, \quad \nonumber\\ &
\qquad\qquad\qquad\qquad\qquad\qquad 
 i=1,2,...,N_S,
 \end{align} where $J_R({\pi})$ and $J_C({\pi})$ are defined as the accumulated reward and central cost while $J_{j(i)}(\pi)$ is the local cost of each agent. Regarding  (11), we aim to maximize the accumulated  average transferring rate of all packets. As for  (12), the accumulated packet loss rate should be kept below $P_L$. Similarly, for (13) and (14), the cumulative energy consumption of each ground station or satellite should remain below respective thresholds $E_G$ and $E_S$.

According to the iterative search method in  policy optimization~\cite{TRPO}~\cite{PPPO}, we reformulate (P2) as:
\begin{align}
\quad \text{(P3)} \quad  & \pi_{k+1} = \underset{\pi}{\arg \max } \underset{\substack{s \sim d^{\pi} \\
\vec{a} \sim \pi}}{{E}}\left[A_{R}^{\pi_{k}}(s, \vec{a})\right] \\
& \text { s.t.} \quad J_{C}\left(\pi_{k}\right)+\frac{1}{1-\gamma} \underset{\substack{s \sim d^{\pi} \\
\vec{a} \sim \pi}}{{E}}\left[A_{C}^{\pi_{k}}(s, \vec{a})\right] \leq P_L,  \\ 
& \quad J_{j}\left(\pi_{k}\right)+\frac{1}{1-\gamma} \underset{\substack{o_j \sim d^{\pi} \\
a_j \sim \pi}}{{E}}\left[A_{j}^{\pi_{k}}(o_j, a_j)\right] \leq E_G, \nonumber \\ \quad & \qquad\qquad\qquad\qquad\qquad\qquad  j=1,2,...,N_G,  \\ 
& \quad J_{i}\left(\pi_{k}\right)+\frac{1}{1-\gamma} \underset{\substack{o_i \sim d^{\pi} \\
a_i \sim \pi}}{{E}}\left[A_{i}^{\pi_{k}}(o_i, a_i)\right] \leq E_S, \quad \nonumber\\ &
\qquad\qquad\qquad\qquad\qquad\qquad 
 i=1,2,...,N_S, \\
& \quad  \underset{s \sim d^{\pi_{k}}}{E}\left[\mathrm{D}_{K L}\left(\pi \| \pi_{k}\right) \right] \leq \delta,
\end{align}
where  $d^{\pi}(s)  = (1-\gamma) \sum_{t  = 0}^{\infty} \gamma^{t} P\left(s_{t} = s \mid \pi\right) $ represents the discounted future state distribution. 
$\pi_{k}$ is the old policy and $\pi_{k+1}$ is the updated policy. $A_{R}^{\pi_{k}}(s, \vec{a})$ is the advantage function and $A_{R}^{\pi_{k}}(s, \vec{a}) = Q_{R}^{\pi_{k}}(s, \vec{a}) - V_{R}^{\pi_{k}}(s, \vec{a})$. $Q_{R}^{\pi_{k}}(s, \vec{a})$ is the action-value function and is defined as $Q_{R}^{\pi_{k}}(s, \vec{a}) = {E}_{ s \sim p, \vec{a} \sim {\pi} }\left[\sum_{t=1}^{W} \gamma^{t} r\left({s}^{t}, {\vec{a}}^{t}\right)\mid {s}^{t} = s, {\vec{a}}^{t} = \vec{a}\right]$. $V_{R}^{\pi_{k}}(s, \vec{a})$ is the value function and is defined as $V_{R}^{\pi_{k}}(s, \vec{a}) = {E}_{ s \sim p, \vec{a} \sim {\pi} }\left[\sum_{t=1}^{W} \gamma^{t} r\left({s}^{t}, {\vec{a}}^{t}\right)\mid {s}^{t} = s\right]$. The inequality $\underset{s \sim d^{\pi_{k}}}{E}\left[\mathrm{D}_{K L}\left(\pi \| \pi_{k}\right) \right] \leq \delta$ limits the update  of each step strategy from $\pi_{k}$ to $\pi_{k+1}$ within a certain range.

Considering the Lagrangian method~\cite{2009Convex}~\cite{Gu2021MultiAgentCP} in solving the optimization problem with constraints, we change the (P3) to a max-min optimisation problem :
\begin{align}
& \text{(P4)    } \max_{\pi }\min_{\lambda_C \geq 0, \lambda_{j(i)} \geq 0} {L}(\pi, \lambda_C, \lambda_{j(i)}) \text{ , where } \nonumber \\
& \quad {L}(\pi, \lambda_C, \lambda_{j(i)}) \triangleq 
\underset{\substack{s \sim d^{\pi} \\
\vec{a} \sim \pi}}{{E}}\left[ 
A_{R}^{\pi_{k}}(s, \vec{a})\right]  \nonumber \\ 
& - \lambda_C \times \Bigg[    \underset{\substack{s \sim d^{\pi} \\
\vec{a} \sim \pi}}{{E}}\left[ 
A_{C}^{\pi_{k}}(s, \vec{a})\right]  + (1-\gamma)(J_{C}\left(\pi_{k}\right) - P_L)  \Bigg]  \nonumber \\ 
& - \sum_{j=1}^{N_G} \lambda_{j} \times \Bigg[    \underset{\substack{o_{j} \sim d^{\pi} \\
a_{j} \sim \pi}}{{E}}\left[ 
A_{j}^{\pi_{k}}(o_{j}, a_{j})\right] + (1-\gamma) \left(J_{j}\left(\pi_{k}\right) - E_{G}\right)  \Bigg] \nonumber \\ 
& - \sum_{i=1}^{N_S} \lambda_{i} \times \Bigg[    \underset{\substack{o_{i} \sim d^{\pi} \\
a_{i} \sim \pi}}{{E}}\left[ 
A_{i}^{\pi_{k}}(o_{i}, a_{i})\right] + (1-\gamma) \left(J_{i}\left(\pi_{k}\right) - E_{S}\right)  \Bigg].
\end{align}
Here, $\lambda_C$,  $\lambda_j$ and $\lambda_i$ 
are Lagrangian multipliers for the global cost constraint function and the  local cost constraint function of each ground station or satellite. Referring to the PPO-Clip~\cite{CPO}, the ratio $r({\theta}_{j(i)}) = \frac{\pi_{\theta_{j(i)}}\left(a_{j(i)}^{t} \mid o_{j(i)}^{t}\right)}{\pi_{\theta_{j(i)}}^{k}\left(a_{j(i)}^{t} \mid o_{j(i)}^{t}\right)}$, where $\pi_{\theta_{j(i)}}^{k}$ is the old routing policy and $\pi_{\theta_{j(i)}}$ is the new policy. The clip operator is used to limit the magnitude of policy updates, ensuring that the ratio between the new and old policies stays within a certain range ($1-\epsilon, 1+\epsilon$), thus avoiding instability~\cite{Schulman2017ProximalPO}.

\subsection{Algorithm Architecture}
To address the (P4) issue, we have developed the CMADR architecture. This architecture consists of an actor network and a local cost critic network for each satellite and ground station, along with a central reward critic network and a central cost critic network for the entire communication system, as depicted in Figure 3. The actor network's role is to generate specific routing schemes, while the local cost critic network evaluates the energy consumption cost value function of each ground station and satellite based on constraints (9) and (10). Furthermore, since the average time delay in objective function (8) and packet loss rate in constraint (7) are influenced by the entire system, the central delay reward critic network and the central packet loss cost critic network are utilized to assess these two components. With the combined feedback from these reward/cost critics, the actors are trained to provide improved routing schemes.

Specifically, each satellite and ground station generates their own local observations ${o}_{{i}}^{t}$ and ${o}_{{j}}^{t}$ at time  slot $t$, which are then transmitted to the respective local actor network and local cost critic network.  Simultaneously, the global state $s^t$ from environment is sent to the central reward critic and the central cost critic to acquire the joint state reward value and joint state cost value. Following the interaction between the satellites/ground stations and the environment, the global reward ${r}^{t}$ and global cost ${c}^{t}$, along with the subsequent local observations ${o}_{{i}}^{t+1}$ and ${o}_{{j}}^{t+1}$ at time slot $t+1$, are obtained from the environment. With transitions stored in a buffer, all critic networks will be trained to learn more accurate value evaluations. In turn, they  collaborate to train the actor network with the objective of reducing the average delay, while  constraining energy consumption and packet loss rate. 

\begin{figure*}[htb]
\centering  
\includegraphics[height=8cm,width=11cm]{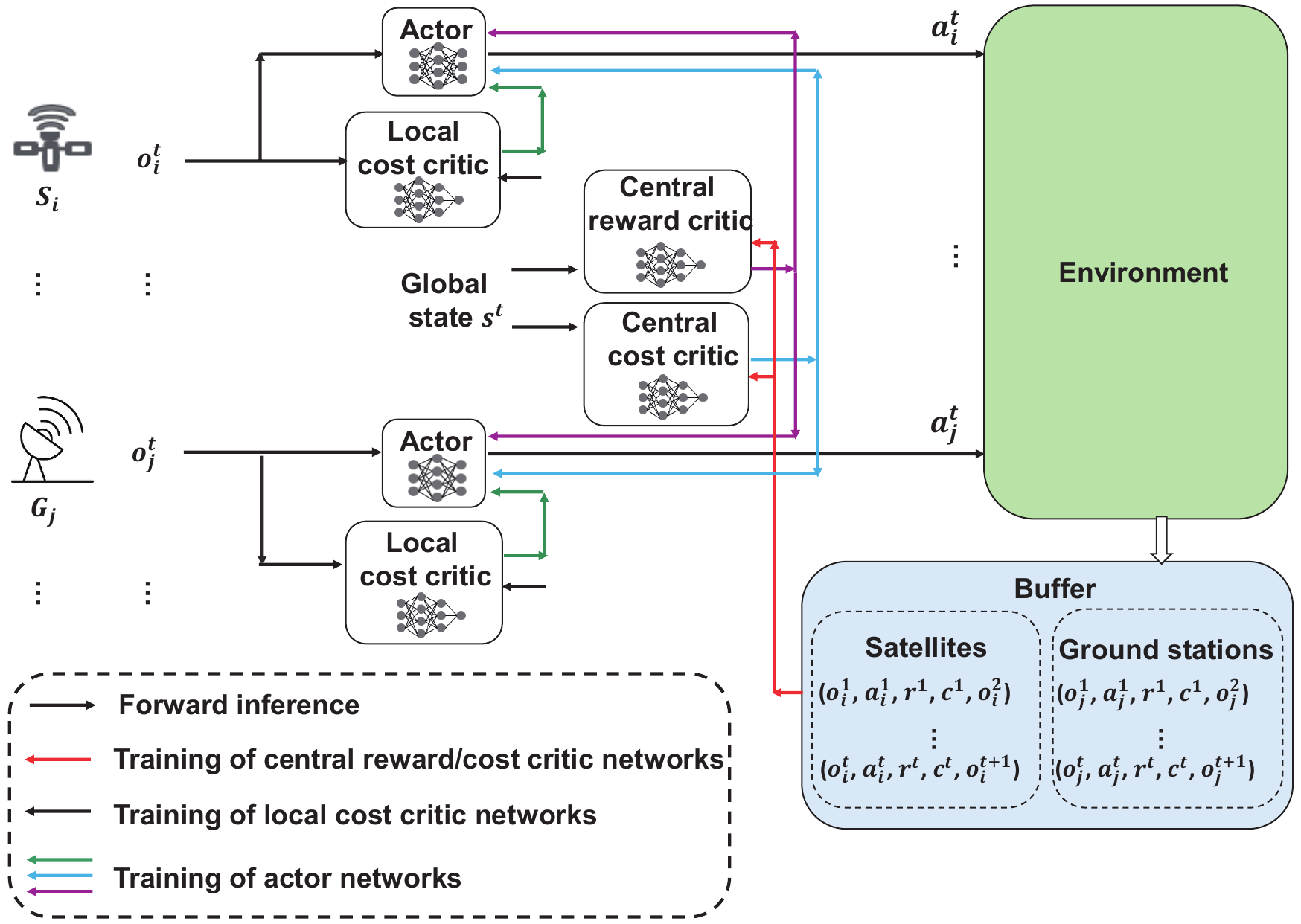}
\vspace{-10pt}
\caption{Architecture of CMADR. At time  slot $t$, each satellite and ground station form their own observations ${o}_{{i}}^{t}$ and ${o}_{{j}}^{t}$  that are transmitted to the local actor and local cost critic. The global state $s$ is sent to the central reward critic and central cost critic to obtain the joint state reward value and joint state cost value for actor network training. Following the interaction, environment provides the global reward ${r}^{t}$ and global cost ${c}^{t}$, as well as the next local observations ${o}_{{i}}^{t+1}$ and ${o}_{{j}}^{t+1}$ at time slot $t+1$. By utilizing the stored transitions, the critic networks will learn more accurate value evaluations and collaborate to train the actor network with the aim of reducing average delay while simultaneously constraining energy consumption and packet loss rate.} 
\vspace{-10pt}
\end{figure*}

\subsection{Training and Executing}
During the training of the communication system, the centralized training with decentralized execution (CTDE)  paradigm~\cite{Kuba2021TrustRP}~\cite{qmix}~\cite{coma} is followed. During the training process, each actor network of the satellite and ground station is trained by the state reward value function and state cost value function from critic networks. However, during the executing process, the actor network relies solely on its own limited observation  to develop a routing strategy without any need for other critic networks to participate.

\textbf{Training Process}.

For each local actor network of ground stations/satellites:
\[Input:\{
{o}_{{j(i)}}^{t}\} \rightarrow  Output:\{ {a}_{{j(i)}}^{t} \}  \]
 $G_j$ and  $S_i$ utilize their own actor network, characterized by parameters $\theta_j$ and $\theta_i$ respectively, to generate a routing scheme ${a}_{{j}}^{t}$ and ${a}_{{i}}^{t}$ correspondingly.

Similarly, for each local cost critic network of ground stations/satellites:
\[Input:\{
{o}_{{j(i)}}^{t}\} \rightarrow  Output:\{ {V}_{{j(i)}}^{t}\} \]
The energy consumption cost values ${V}_{{j}}^{t}$ and ${V}_{{i}}^{t}$are generated by local cost critic networks, utilizing parameters $\phi_j$ and $\phi_i$, respectively, relying on observations ${o}_{{j}}^{t}$ and ${o}_{{i}}^{t}$.

For the central reward/cost critic network:
\[Input:\{
{s}^{t}\} \rightarrow  Output:\{ {V}_{R}^{t}, {V}_{C}^{t}\} \]
The central reward/cost critic network with parameters $\phi_R$ and $\phi_C$ generates the   reward value  ${V}_{R}^{t}$ and the cost value ${V}_{C}^{t}$ based on the global state $s$.

The gradient of updating each actor network can be represented as follows:
\begin{align}
& \Delta_{\theta_{j(i)}}= 
\nabla_{\theta_{j(i)}}   ( L_R - \lambda_C \times L_C 
 - \lambda_{j(i)} \times  L_{j(i)}   ) \text{ , where }\nonumber \\  &  L_R  =  \underset{\substack{s \sim d^{\pi} \\
\vec{a} \sim \pi}}{{E}}\Big[  
min \Big \{ r({\theta}_{j(i)})  A_{R}^{\pi_{k}}(s, \vec{a}), 
clip(r({\theta}_{j(i)}), 
 1 \pm \epsilon ) A_{R}^{\pi_{k}}(s, \vec{a})      \Big \}           \Big] ,  \nonumber \\  &
 L_C  = 
 \underset{\substack{s \sim d^{\pi} \\
\vec{a} \sim \pi}}{{E}}\Big[  
min \Big \{ r({\theta}_{j(i)})  A_{C}^{\pi_{k}}(s, \vec{a}), \nonumber \\  &
clip(r({\theta}_{j(i)}), 
 1\pm \epsilon ) A_{C}^{\pi_{k}}(s, \vec{a})      \Big \}             + (1-\gamma)(J_{C}\left(\pi_{k}\right) - P_L) \Big],
\nonumber \\ &  L_{j(i)}  = \underset{\substack{o_{j(i)} \sim d^{\pi} \\
a_{j(i)} \sim \pi}}{{E}}\Big[  
min \Big \{ r({\theta}_{j(i)})  A_{j(i)}^{\pi_{k}}(o_{j(i)}, a_{j(i)}), \nonumber \\  &
clip(r({\theta}_{j(i)}), 
 1\pm \epsilon ) A_{j(i)}^{\pi_{k}}(o_{j(i)}, a_{j(i)})      \Big \}   + (1-\gamma)(J_{C}\left(\pi_{k}\right) - P_L) \Big].  
\end{align}
We update parameters of each actor network as follows:
\begin{equation}
\theta_{j(i)}^{k+1} = \theta_{j(i)}^{k} +\alpha_{\theta} \times  \Delta_{\theta_{j(i)}},
\end{equation}
where $\alpha_{\theta}$ is the learning rate of actor networks. We calculate the update rate of  $\lambda_C$  as:
\begin{align}
\Delta_{{\lambda}_{C}} = \Bigg[    \underset{\substack{s \sim d^{\pi} \\
\vec{a} \sim \pi}}{{E}}\left[ 
A_{C}^{\pi_{k}}(s, \vec{a})\right]  + (1-\gamma)(J_{C}\left(\pi_{k}\right) - P_L)  \Bigg],
\end{align}
and calculate the update rate of  $\lambda_j$ and $\lambda_i$  as:
\begin{align}
&\Delta_{{\lambda}_{j}} = \Bigg[    \underset{\substack{o_{j} \sim d^{\pi} \\
a_{j} \sim \pi}}{{E}}\left[ 
A_{j}^{\pi_{k}}(o_{j}, a_{j})\right] + (1-\gamma) \left(J_{j}\left(\pi_{k}\right) - E_{G}\right)  \Bigg] \text {, and }  \\   &
\Delta_{{\lambda}_{i}} = \Bigg[    \underset{\substack{o_{i} \sim d^{\pi} \\
a_{i} \sim \pi}}{{E}}\left[ 
A_{i}^{\pi_{k}}(o_{i}, a_{i})\right] + (1-\gamma) \left(J_{i}\left(\pi_{k}\right) - E_{S}\right)  \Bigg].
\end{align}
With $\Delta_{{\lambda}_{C}}$, $\Delta_{{\lambda}_{j}}$ and $\Delta_{{\lambda}_{i}}$, we update Lagrangian multipliers as follows~\cite{Gu2021MultiAgentCP}:
\begin{align}
& \lambda_C^{k+1} = ReLU({\lambda}_C^{k}+ {\alpha}_{\lambda}\times \Delta_{{\lambda}_{C}}) \text {, and }  \\   & 
\lambda_{j(i)}^{k+1} = ReLU({\lambda}_{j(i)}^{k}+ {\alpha}_{\lambda}\times \Delta_{{\lambda}_{{j(i)}}}),
\end{align}
where ${\alpha}_{\lambda}$ is the step size of  multipliers and the rectified linear unit (ReLU) function~\cite{Gu2021MultiAgentCP} guarantees that the multipliers remain positive.
Finally, we calculate the gradient of central reward/cost critic network by commonly employed  mean squared error (MSE)~\cite{PPPO} as:
\begin{align}
&\Delta_{\phi_{R}}=\nabla_{\theta_{R}}\sum_{t=1}^{W}{({V}_{R}^{t} - R^t)}^2
\text {, and } \nonumber \\   & 
\Delta_{\phi_{C}}=\nabla_{\theta_{C}}\sum_{t=1}^{W}{({V}_{C}^{t} - C^t)}^2,
\end{align}
where $R^t$ and $C^t$ are calculated following the reward-to-go~\cite{Kuba2021TrustRP} that $R^t = \sum_{l=0}^{L} \gamma^l \times {r}^{t+l}$ and $C^t = \sum_{l=0}^{L} \gamma^l \times {c}^{t+l}$. Similarly, the gradient of  local cost critic networks is:
\begin{equation}
\Delta_{\phi_{j(i)}}=\nabla_{\theta_{j(i)}}\sum_{t=1}^{W}{({V}_{{j(i)}}^{t} - {C}_{j(i)}^{t})}^2,
\end{equation}
where ${C}_{j(i)}^{t}) = \sum_{l=0}^{L} \gamma^l \times {c}_{j(i)}^{t+l}$. With $\Delta_{\phi_{R}}$, 
$\Delta_{\phi_{C}}$ and 
$\Delta_{\phi_{j(i)}}$, we update parameters of all critic networks as follows:
\begin{align}
&\ \ \ \phi_{R}^{k+1} = \phi_{R}^{k} +\alpha_{\phi} \times  \Delta_{\phi_{R}} \text{, }
\phi_{C}^{k+1} = \phi_{C}^{k} +\alpha_{\phi} \times  \Delta_{\phi_{C}} ,  \\ & \  \  \ \  \
\phi_{j(i)}^{k+1} = \phi_{j(i)}^{k} +\alpha_{\phi} \times  \Delta_{\phi_{j(i)}},
\end{align}
where $\alpha_{\phi}$ is the learning rate of critic networks.

\textbf{Executing Process (only actor)}.
In a real integrated satellite-ground  service scenario, the interaction between the environment and all satellites and ground stations is highly complex. Due to the frequent information exchange between space and the ground, collecting data and performing gradient calculations are expensive. 

To address this problem, a more suitable approach is to simulate this communication system and gather data before launching the satellites, training the hardware that carries the neural network. Taking into account the limited capacity of the satellite, once the training is completed, the lightweight neural network can be deployed on a GPU. This enables it to solely perform forward inference in real satellite communication scenarios, eliminating the need for backward training and resulting in significant cost savings. During the execution phase, each satellite and ground station only require their respective actor network without any additional critic networks, which makes the overall consumption acceptable.  

We summarize details of CMADR in Algorithm 1.

\begin{algorithm}
\caption { CMADR }
\textbf{Initializing phase:} 

Initialize batch size $B$, number of iterations $K$,  number of steps per episode $W$, replay buffer  $\mathcal{B}$;\\
Initialize  actor networks and local critic networks as $\left\{\theta_{j(i)}^{0}, \phi_{j(i)}^{0},{\forall j, i} \right\}$;\\
Initialize the central reward critic network and the central cost critic network as $\left\{\phi^{0}_{R},\phi^{0}_{C}\right\}$; \\ 
Initialize Lagrangian multipliers $\lambda_C$,  $\lambda_j$ and $\lambda_i$;\\
\For(){$k=1, \ldots, K$}{
\For(){$t=1, \ldots, W$}{
Interact with the environment by  the joint policy ${\pi} = \{ {\pi}_{{\theta}_{j}}, {\pi}_{{\theta}_{i}},{\forall j, i}\}$ and store transitions  $({o}_{j}^{t},{a}_{j}^{t}, {r}^{t}, {c}^{t}, {o}_{j}^{t+1})$ and $({o}_{i}^{t},{a}_{i}^{t}, {r}^{t}, {c}^{t}, {o}_{i}^{t+1})$  for ground stations and satellites in buffer $\mathcal{B}$; \\
}
Sample a batch of $B$ transitions from the buffer $\mathcal{B}$;\\
\For(){$j=1,2,...,N_G, i=1,2,...,N_S$}{
Update actor parameters $\theta_{j(i)}$ by equation (21) and (22);\\
Update  Lagrangian multipliers $\lambda_C$ of the central critic and $\lambda_{j(i)}$ of local critics by equation (23) and (26), as well as  equation (24), (25), and (27);

}
Update parameters of the  central reward/critic network and local critic networks $\phi_{R}$,  $\phi_{C}$, and $\phi_{j(i)}$ by equation (28), (30), (29) and (31);\\
}

\end{algorithm}

\section{SIMULATION}

In this section, we evaluate the effectiveness of CMADR in ISTN by comparing it with other baseline approaches, focusing on two constellations, Telesat and OneWeb. Additionally, we perform an ablation study to showcase the contributions of the global critic network and local critic networks in addressing the packet loss constraint and energy consumption constraints. Furthermore, we conduct experiments to adjust various constraint thresholds, illustrating that CMADR is capable of satisfying diverse constraint conditions.

\subsection{Experimental Settings}

We have conducted extensive research on the constellation information of Telesat and OneWeb by analyzing the filings submitted to the Federal Communications Commission (FCC)~\cite{US}. Telesat operates at an orbital height of 1015km, with 13 satellites deployed in each orbit. With a total of 27 orbits, the Telesat constellation comprises a total of 351 satellites. On the other hand, OneWeb operates at an orbital height of 1200km, with 40 satellites deployed in each orbit. The OneWeb constellation consists of 18 orbits, resulting in a total number of 720 satellites.  Through referencing the specific orbital parameters of these constellations, including altitude, inclination, eccentricity, and minimum elevation angle, we have constructed a comprehensive satellite constellation model. This model allows us to determine the location information of the satellites at any given time. Additionally, we search positions of eight ground stations across different continents, which ensures that users from around the world can establish communication with one another by connecting to these ground stations and utilizing the satellite communication network. All accessible satellites for each ground station and connectable ground stations for each satellite are calculated depending on positions of satellites and ground stations.

To show the effectiveness of CMADR, we  utilize the following comparison algorithms: \textbf{Dijkstra} ~\cite{553679}  makes all satellites constructing the routing tables by Dijkstra algorithm and updates routing tables at each time slot. Moreover, the following three algorithms are about the connection relationships between satellites and ground stations, which are important in the whole communication system: \textbf{LRST}~\cite{7275422} connects to the satellite with the longest remaining service time until it moves out of the transmission range. \textbf{NSD}~\cite{NSD} connects to the nearest satellite until it goes out of the transmission range. \textbf{CSGI}~\cite{9796886} (Coordinated Satellite-Ground Interconnecting)  selects the satellite closest to the center position of each ground station from the satellite cluster that is accessible to it for connectivity. To assess the efficacy of CMADR in comparison to similar AI algorithms, we have chosen two state-of-the-art (SOTA) multi-agent reinforcement learning algorithms that have demonstrated excellent performance. \textbf{HMF}~\cite{10012942}  incorporates the concept of mean field theory to characterize the interplay between agents and their neighbors, and it adopts this notion to develop the conventional deep Q-learning methodology for training individual satellites and ground stations within the ISTN. \textbf{MACPO}~\cite{GU2023103905} integrates central critics and local critics, explores viable policies within a trust region, improves overall performance, and guarantees constraint adherence through the solution of an approximate quadratic optimization problem.

Environmental configurations and algorithm parameters  are shown in Table II.

\begin{table}[htbp]
	\centering
	\caption{ENVIRONMENTAL CONFIGURATIONS AND CMADR PARAMETERS.}
	\begin{tabular}{|c|c|}
    \hline	 
    \textbf{PARAMETERS}&\textbf{VALUES} \\ 
    \hline	 
 
	    \textbf{SCENARIO CONFIGURATIONS}&\\
		 Total period of time, ${T}$&  120min \\
		 Total time slots, ${W}$& 120 \\
     
        Transmission power $ {P}_{R}, {P}_{O}$& 5w, 5w \\

        Bandwidth, ${B}_{R}, {B}_{O}$& 500MHz, 500MHz \\

        Packet size, ${L}_{P}$&  64kbits\\ 
        
        Signal propagation velocity, $H$& $3 \times 10^8$m/s\\

		Central carrier
        frequency, ${f}_{c}$&28GHz\\

		 Transmitting  antenna gain, ${G}_{T}$& 45dBi\\ 
		Receiving antenna gain, ${G}_{R}$ & 30dBi\\	
         Noise spectral density, ${N}_{0}$ & -174dBm/Hz \\
         ASNR,  $\gamma_{{F}}$  & 25dB ($\alpha$ = 0.1) \\
         Visibility, $L$ & 15Km \\
         Reference SNR , $\gamma_{O}$ & 1e9 \\
         Wavelength , $\lambda$  & 1550nm\\
         
         \hline

		  \textbf{ CMADR PARAMETERS} &\\
		  Buffer capacity, ${B_M}$&10000 \\
		Batch size, ${B}$ &128 \\		
		Learning rate, $\alpha_{\theta}$, $\alpha_{\phi}$, ${\alpha}_{\lambda}$ &1e-4, 1e-4, 1e-3  \\		
        Discount factor, ${\gamma}$&0.96 \\
        PPO-Clip, $\epsilon$&0.2 \\
		Total training episodes, $K$&300 \\
		Optimizer&Adam \\
		\hline

	\end{tabular}

\end{table}
\vspace{-10pt}

\subsection{Performance Comparison}

We set the energy consumption threshold to 10KJ for each satellite and ground station, and the packet loss rate threshold to 1\%. We demonstrate the performance of our CMADR algorithm by presenting the accumulated reward, the accumulated average packet delay, the accumulated energy consumption of all satellites and ground stations, and the accumulated packet loss rate for each episode. A total of 300 episodes were conducted, and the results are depicted in Figure 4 (a), (b), (c), and (d) respectively, representing the Telesat  constellation. The accumulated reward in Figure 4 (a) refers to the sum of rewards obtained in each time slot, with a total of 120 slots in each episode. As the training progresses, it becomes evident that the reward curve experiences rapid growth, while the accumulated energy consumption and packet loss rate  curves in Figure 4 (c) and (d) converge and remain within their respective threshold limits. This indicates that the dynamic routing policy has been optimized to achieve efficient packet forwarding while satisfying the constraints on energy consumption and packet loss rate.

Furthermore, by observing Figure 4 (b), (c), and (d), it becomes apparent that the performance of other algorithms is inferior to our CMADR algorithm. The Dijkstra algorithm always selects the shortest path for each time slot but fails to consider alternative routes when satellite-ground connections switch, resulting in relatively higher packet delays. Moreover, it neglects the constraints on energy consumption and packet loss rate, causing its curves to exceed the specified thresholds. On the other hand, although the LRST, NSD, and CSGI algorithms consider the significance of the satellite-ground connection scheme, they still fall short in accounting for all constraints. This is evident from their curves failing to meet the required constraint specifications. While both the HMF and MACPO can achieve reward convergence, their performance in terms of average delay and meeting the constraints of energy consumption and packet loss rate is suboptimal. This indicates that CMADR algorithm surpasses them in overall performance.

Similar observations can be made for the OneWeb constellation, as depicted in Figure 5 (a), (b), (c), and (d). CMADR outperforms the other comparative algorithms in terms of average packet delay and meeting the various constraint requirements.

\begin{figure*}  
\vspace{-10pt}
\subfigure[Reward]{
\begin{minipage}[t]{0.25\linewidth}
\centering
\includegraphics[height=3.5cm,width=4cm]{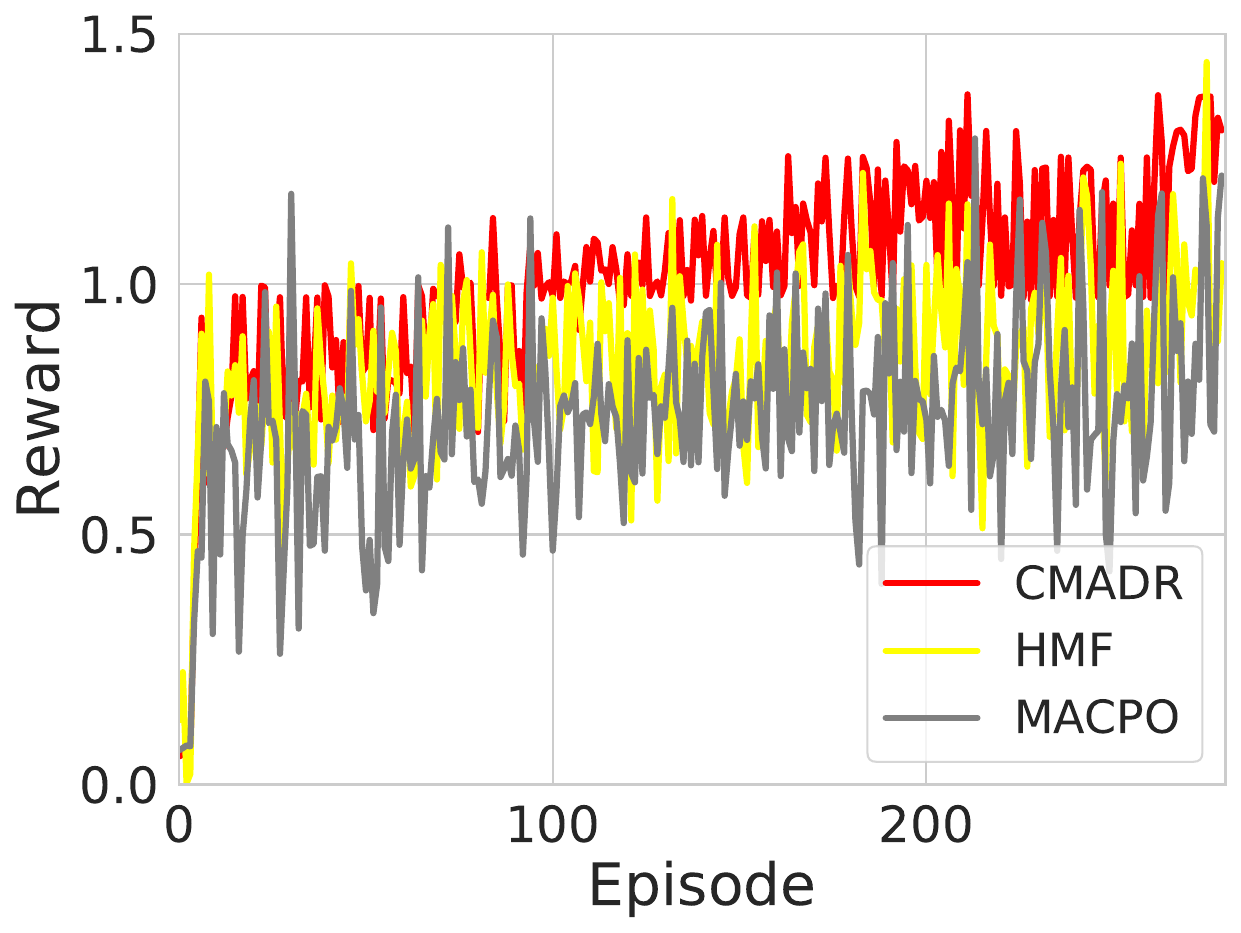}
\end{minipage}%
}%
\subfigure[Average delay]{
\begin{minipage}[t]{0.25\linewidth}
\centering
\includegraphics[height=3.5cm,width=4cm]{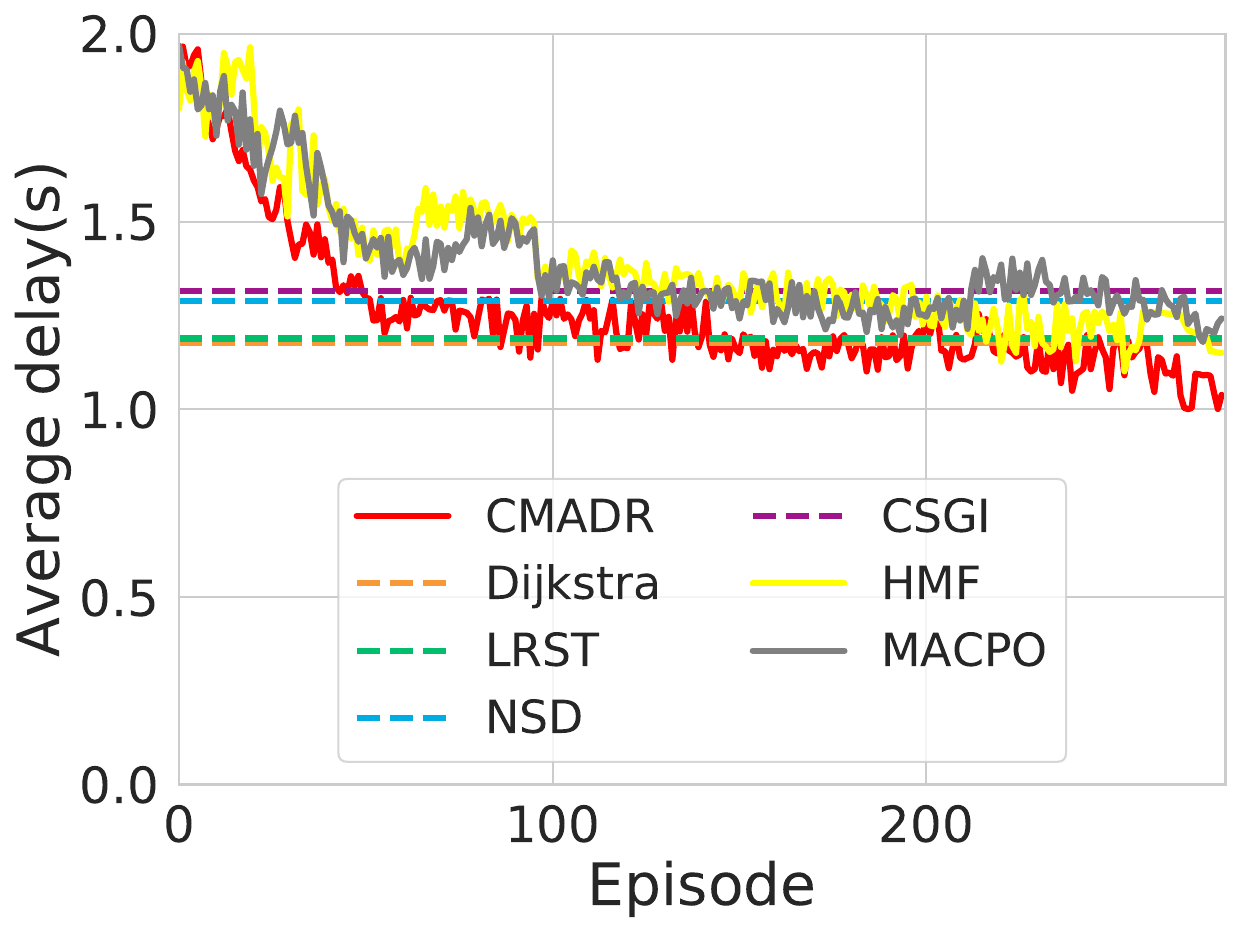}
\end{minipage}%
}%
\subfigure[Energy consumption]{
\begin{minipage}[t]{0.25\linewidth}
\centering
\includegraphics[height=3.5cm,width=4cm]{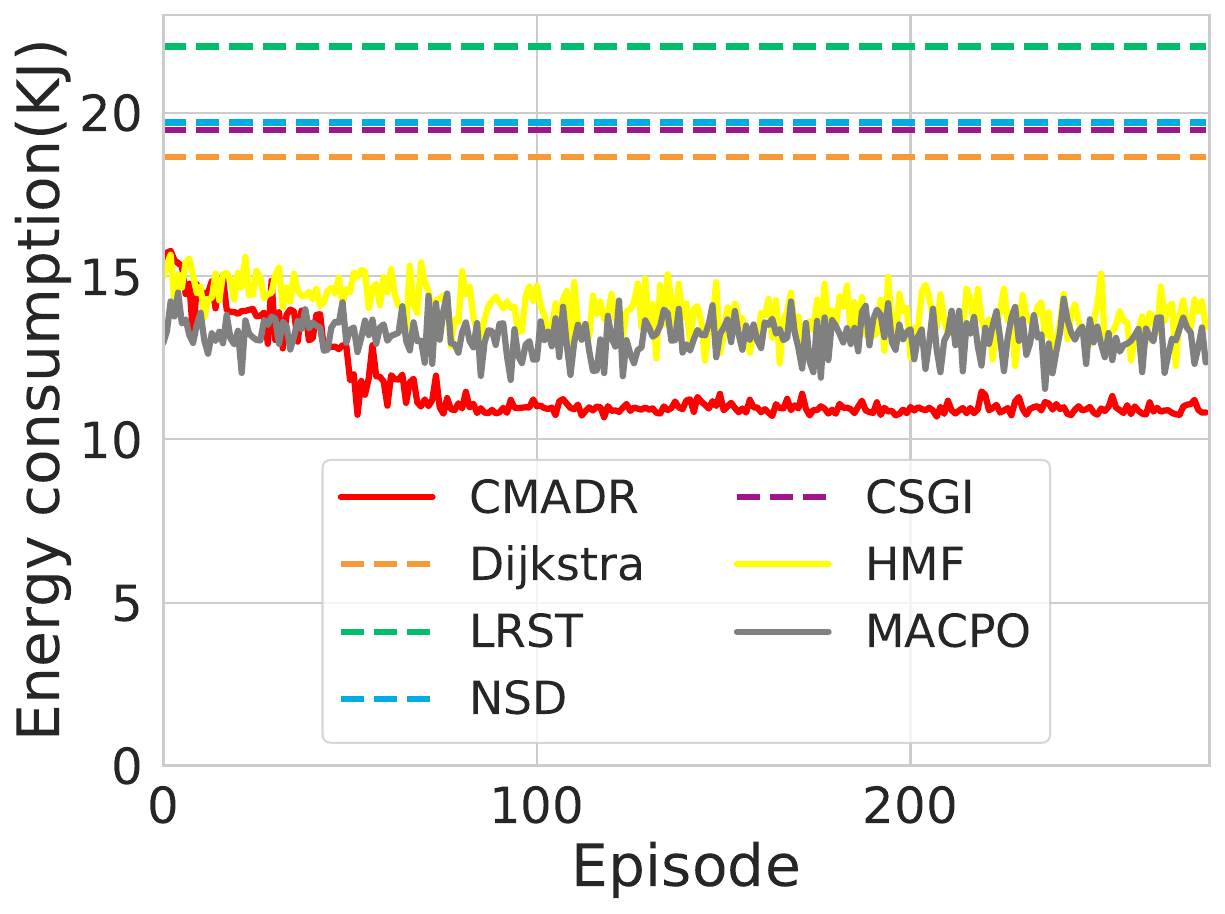}
\end{minipage}%
}%
\subfigure[Packet loss rate]{
\begin{minipage}[t]{0.25\linewidth}
\centering
\includegraphics[height=3.5cm,width=4cm]{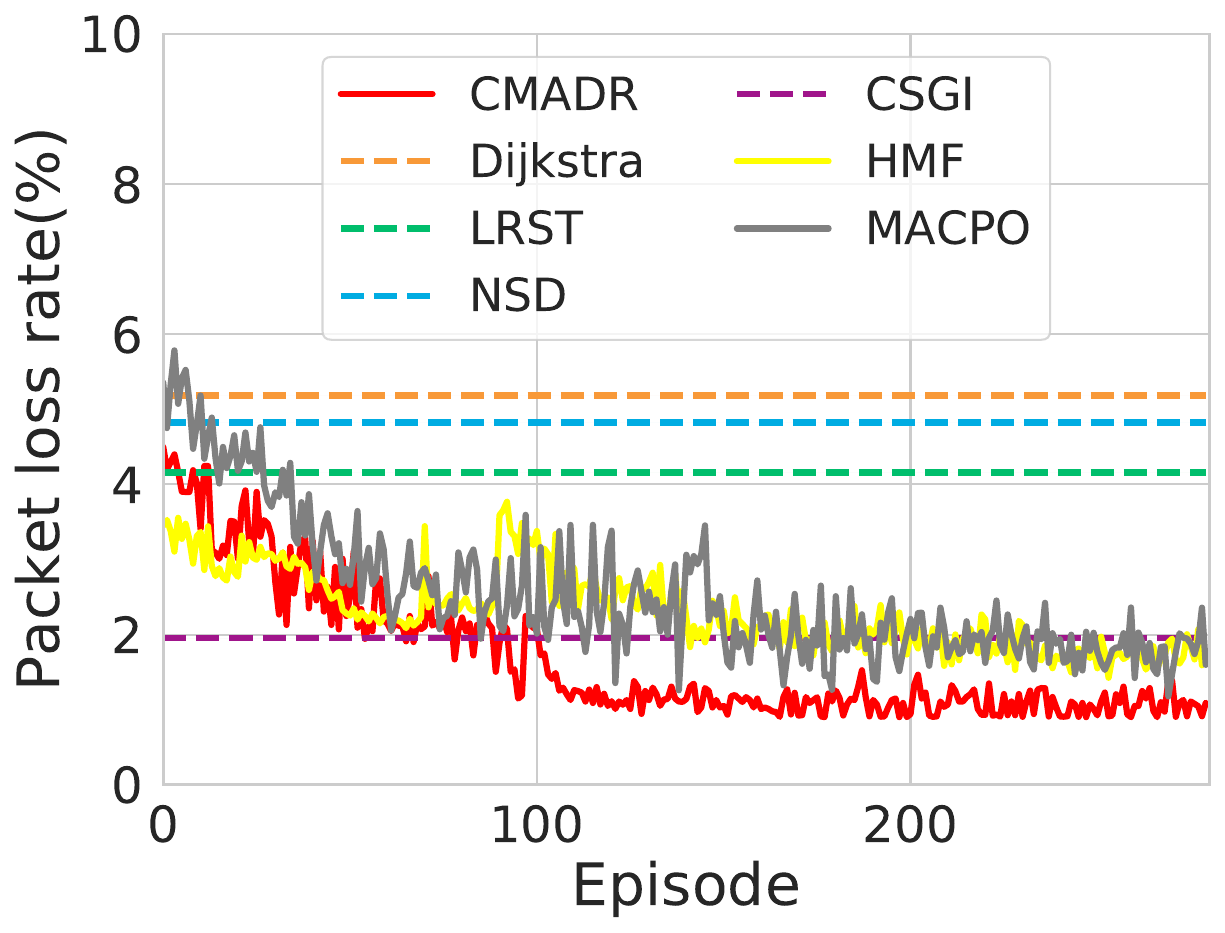}
\end{minipage}%
}%
\caption{(a) The accumulated reward, (b) the average delay, (c) the energy consumption and (d) the packet loss rate  within each episode for 300 episodes based on Telesat. The reward curve of CMADR is growing while the average delay curve is declining. Meanwhile, the   energy consumption and the packet loss rate curves are declining, ultimately satisfying their respective constraints. CMADR performs the best among all algorithms.}
\vspace{-0.3cm}
\end{figure*}

\begin{figure*}  
\vspace{-3pt}
\subfigure[Reward]{
\begin{minipage}[t]{0.25\linewidth}
\centering
\includegraphics[height=3.5cm,width=4cm]{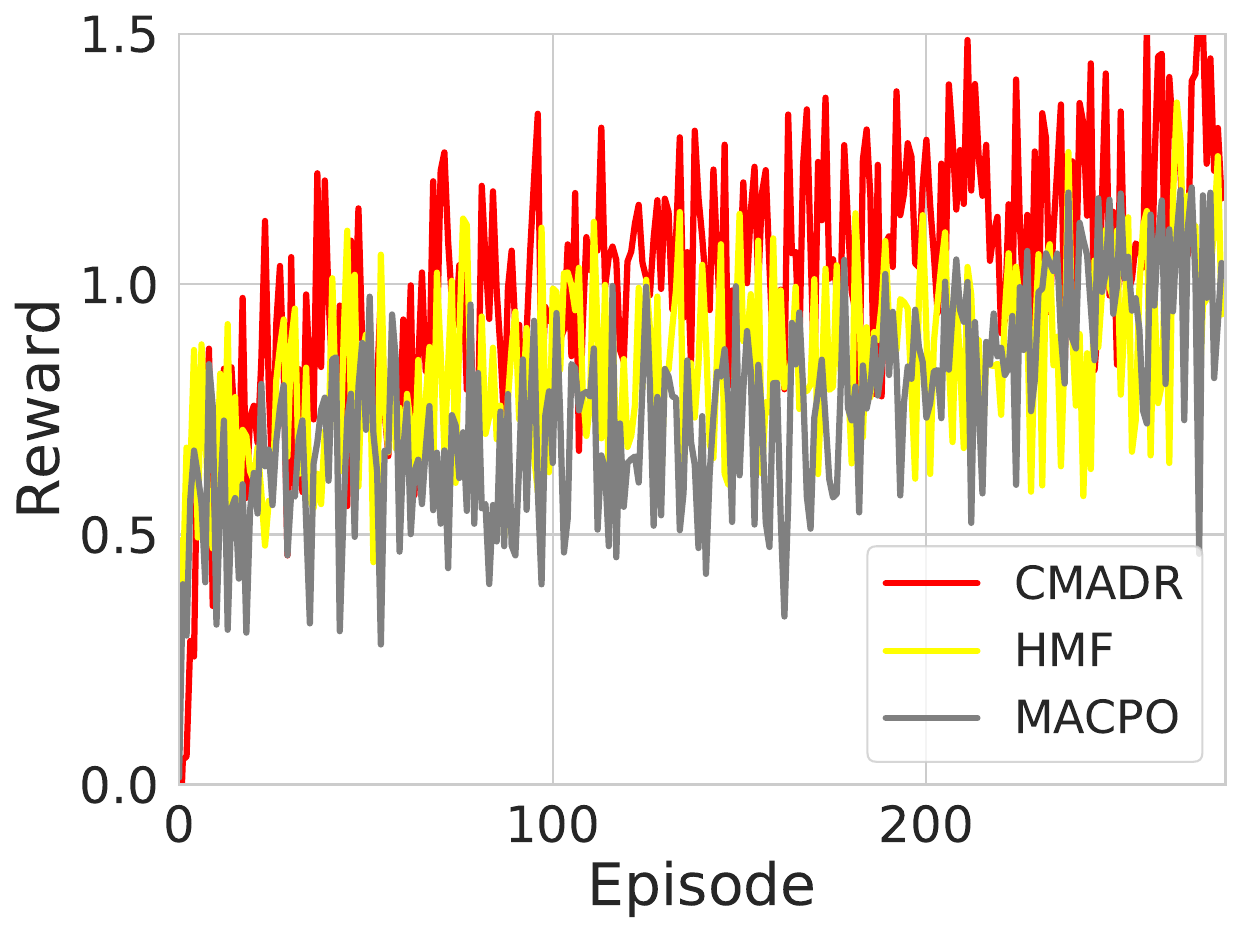}
\end{minipage}%
}%
\subfigure[Average delay]{
\begin{minipage}[t]{0.25\linewidth}
\centering
\includegraphics[height=3.5cm,width=4cm]{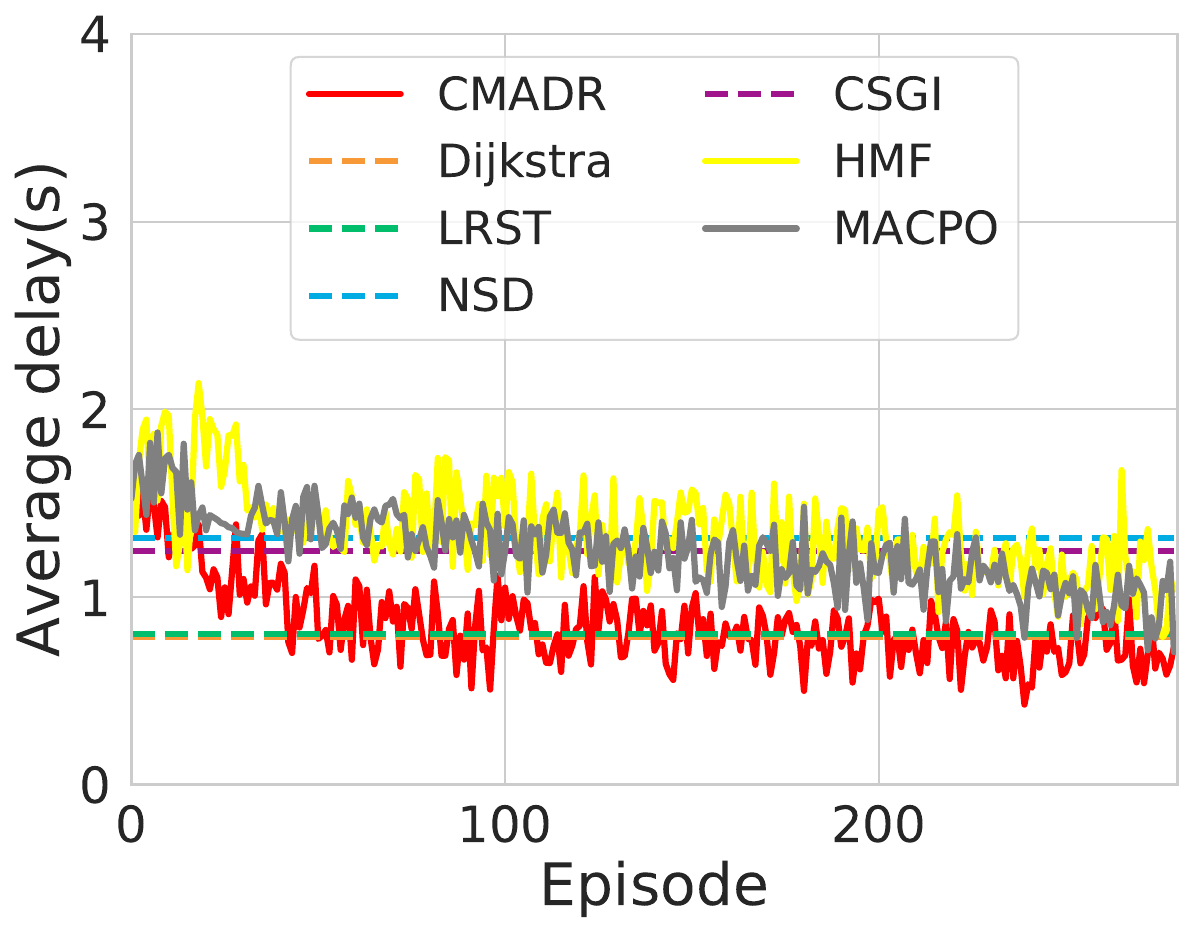}
\end{minipage}%
}%
\subfigure[Energy consumption]{
\begin{minipage}[t]{0.25\linewidth}
\centering
\includegraphics[height=3.5cm,width=4cm]{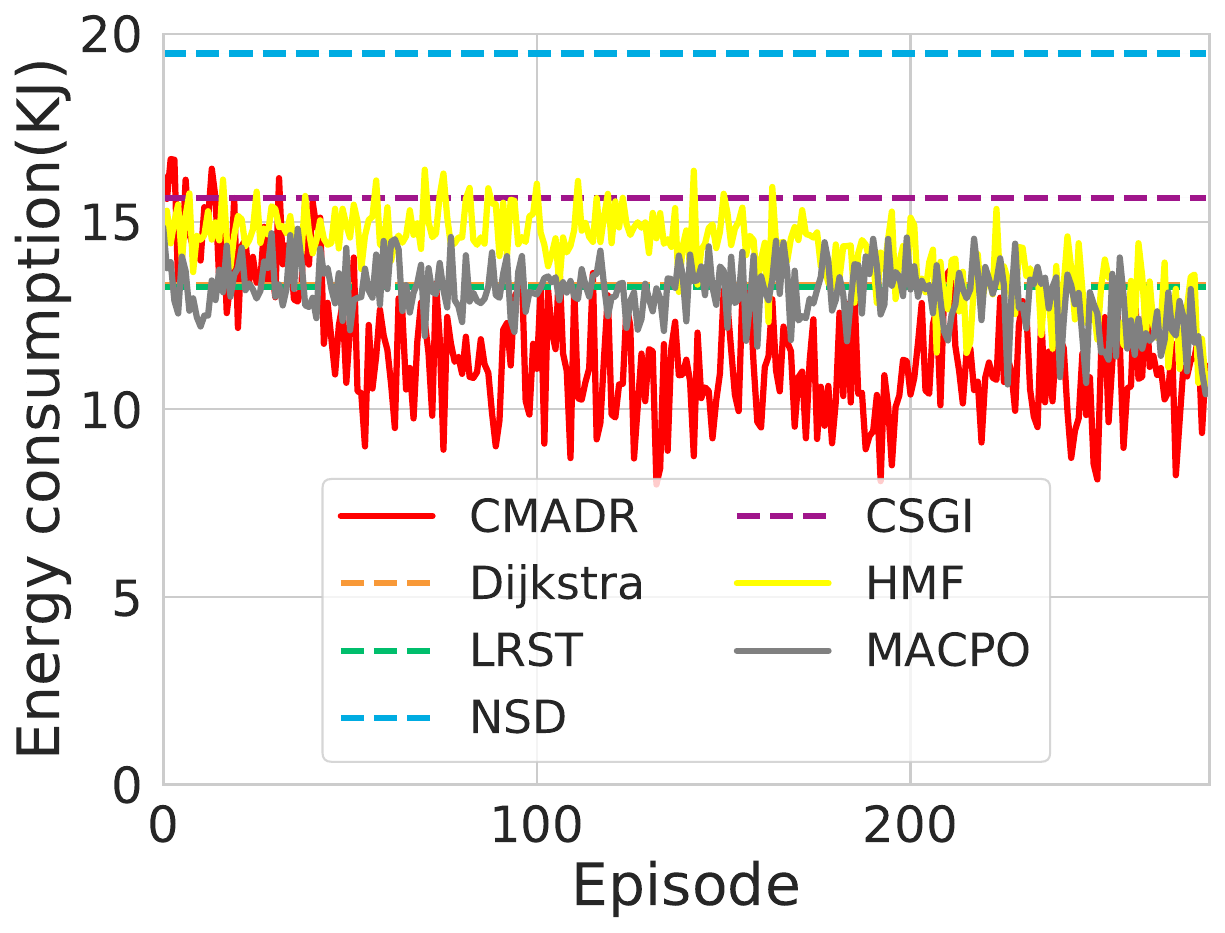}
\end{minipage}%
}%
\subfigure[Packet loss rate]{
\begin{minipage}[t]{0.25\linewidth}
\centering
\includegraphics[height=3.5cm,width=4cm]{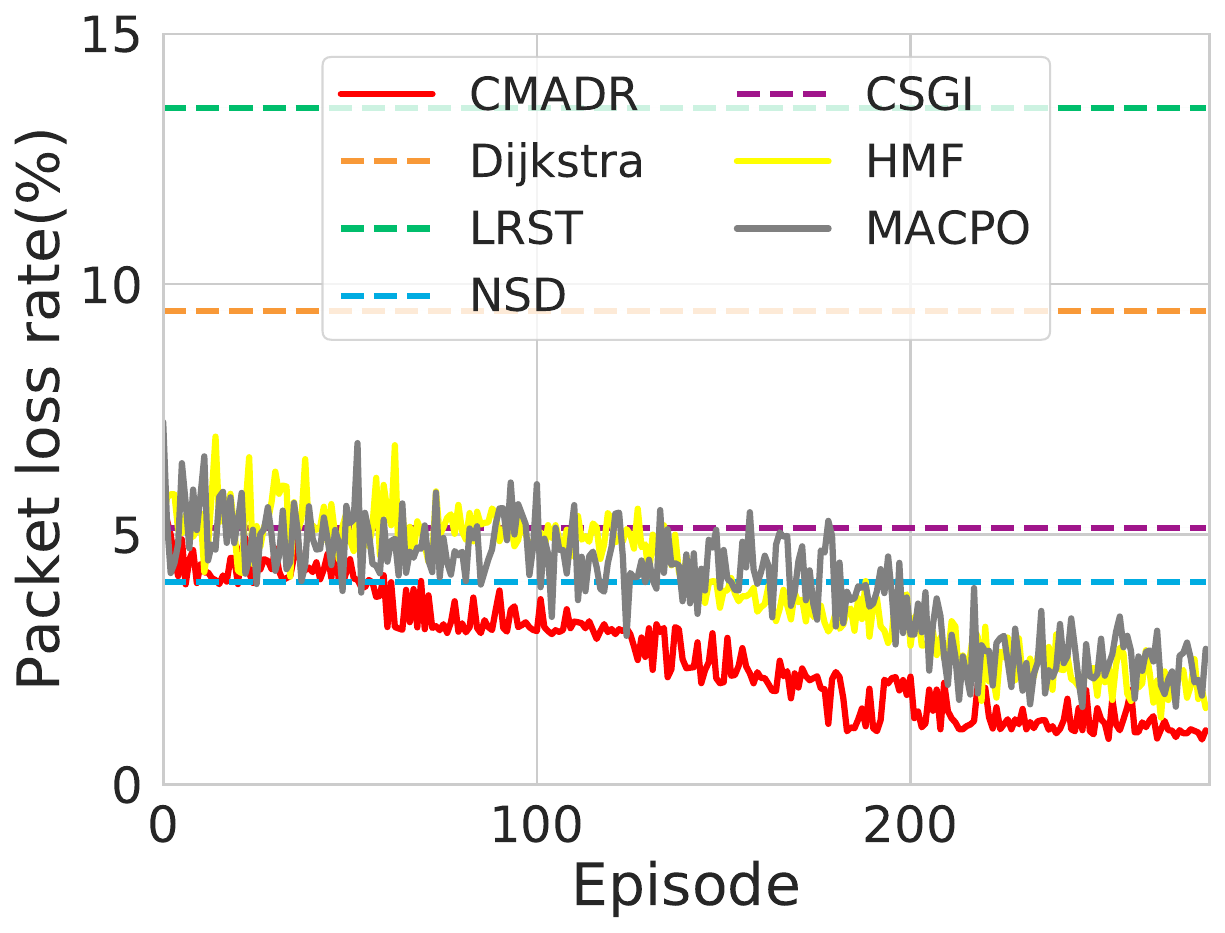}
\end{minipage}%
}%
\caption{(a) The accumulated reward, (b) the average delay, (c) the energy consumption  and (d) the packet loss rate  within each episode for 300 episodes based on OneWeb. Similarly, the reward curve of CMADR shows steady growth, while the average delay curve exhibits a decline. Additionally, both the energy consumption  and the packet loss rate curves are on a downward trend. Overall, CMADR outperforms all other algorithms.}
\vspace{-0.3cm}
\end{figure*}

\begin{figure*}[htbp]  
\vspace{-10pt}
\subfigure[Average delay]{
\begin{minipage}[t]{0.33\linewidth}
\centering
\includegraphics[height=4cm,width=4.5cm]{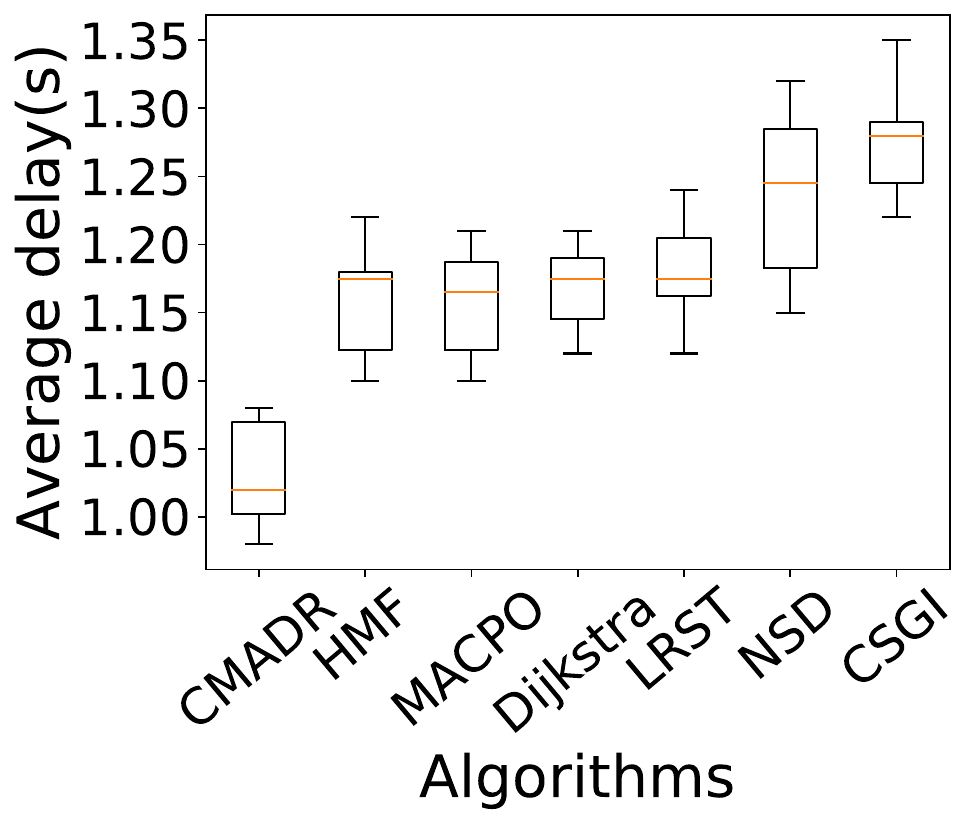}
\end{minipage}%
}%
\subfigure[Energy consumption]{
\begin{minipage}[t]{0.33\linewidth}
\centering
\includegraphics[height=4.3cm,width=4.5cm]{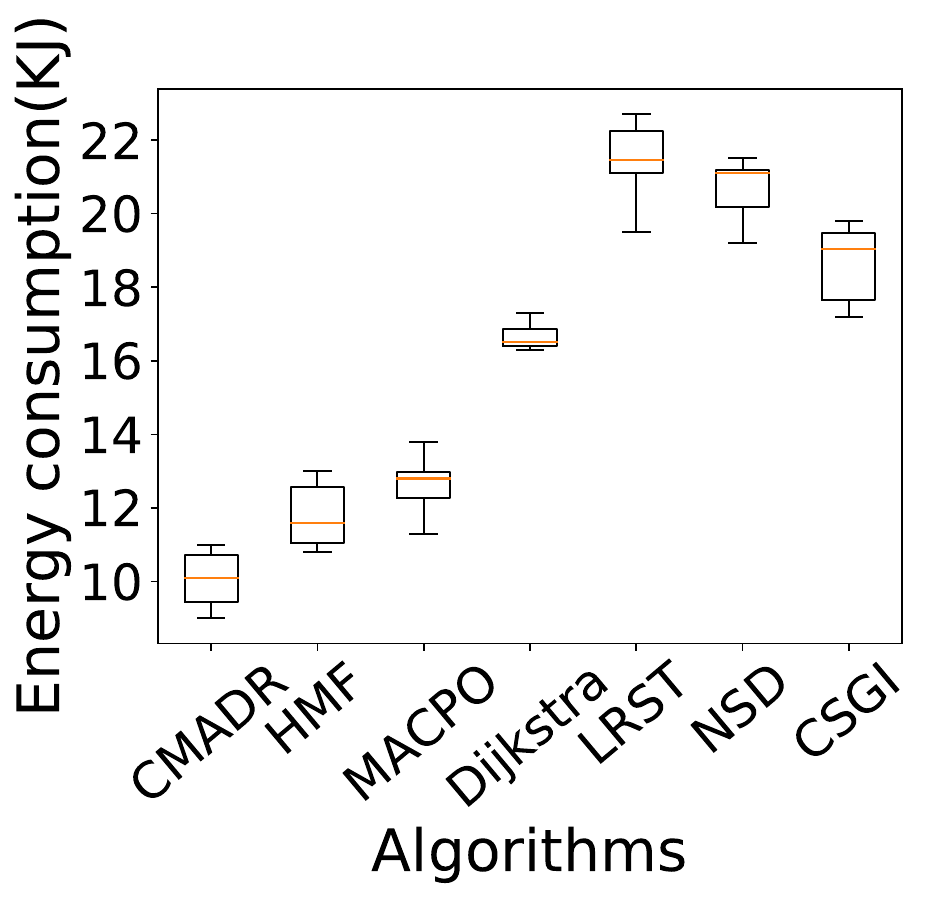}
\end{minipage}%
}%
\subfigure[Packet loss rate]{
\begin{minipage}[t]{0.33\linewidth}
\centering
\includegraphics[height=4cm,width=4.5cm]{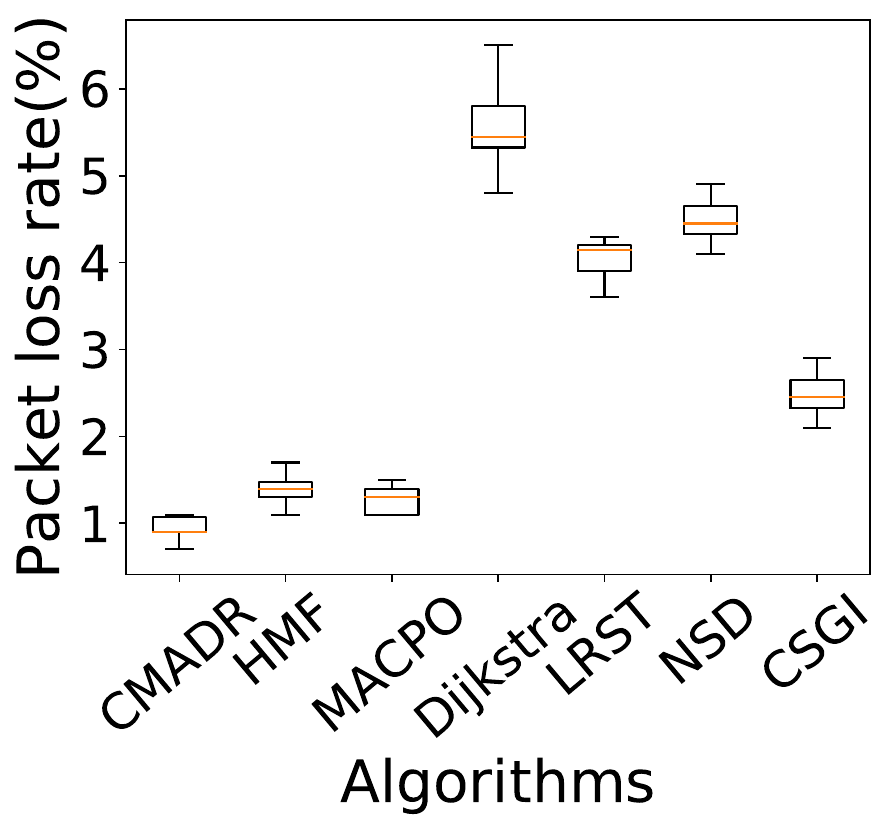}
\end{minipage}%
}%
\caption{(a) The average delay, (b) the energy consumption and (c) the packet loss rate for all algorithms were evaluated in 100 tests based on Telesat.   By exclusively utilizing the trained actor network without any interactions with critics, the CMADR algorithm demonstrates the best performance compared with other methods.}
\vspace{-0.3cm}
\end{figure*}

The results of  AI algorithms shown in Figure 4 and Figure 5 are obtained during the training phase, involving information exchange with critics. In order to demonstrate the effectiveness of the well-trained network in different testing environments that are distinct from the training environment, we conducted a total of 100 tests for all algorithms. These tests involved simulating satellite failures and ground station additions by randomly reducing ten satellites and increasing two ground stations each time. By solely utilizing the trained actor network without any exchanges with critics, the test results in Figure 6 and Figure 7 confirmed the strong performance of our proposed CMADR algorithm. Due to potential satellite failures and the inevitable introduction of new ground stations, the network topology is bound to undergo changes, rendering various methods unstable when confronted with different environments. In contrast to alternative algorithms, our CMADR exhibits relatively stable performance and, on average, outperforms all other methods.

\begin{figure*}[ht]  
\vspace{-10pt}
\subfigure[Average delay]{
\begin{minipage}[t]{0.33\linewidth}
\centering
\includegraphics[height=4cm,width=4.5cm]{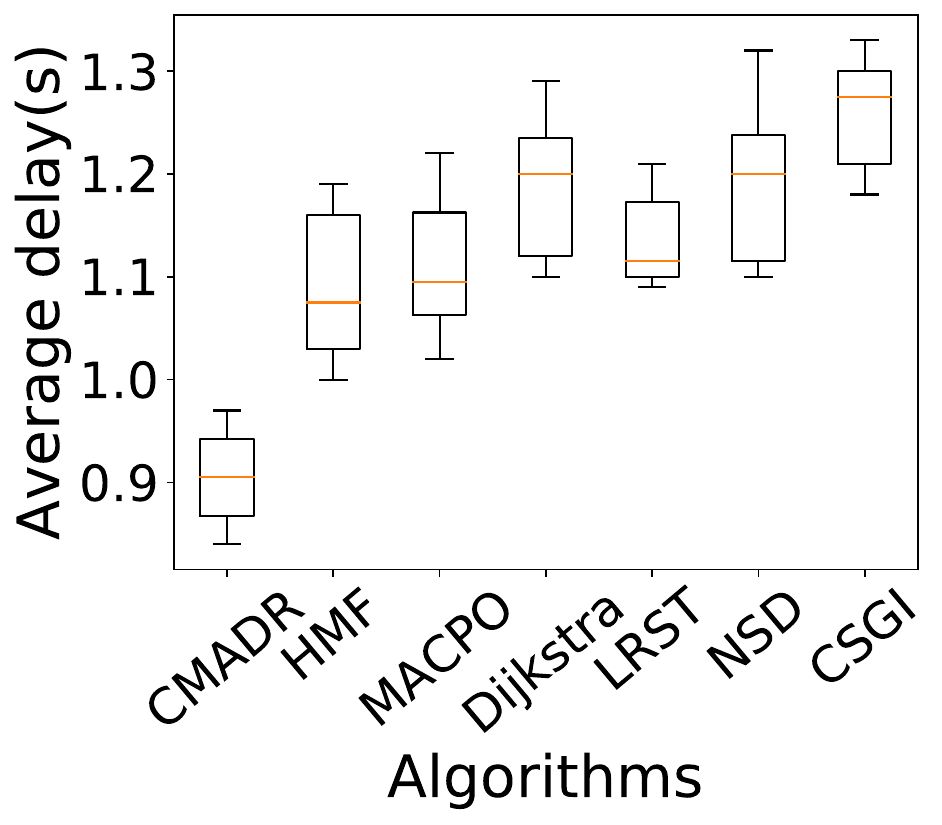}
\end{minipage}%
}%
\subfigure[Energy consumption]{
\begin{minipage}[t]{0.33\linewidth}
\centering
\includegraphics[height=4.3cm,width=4.5cm]{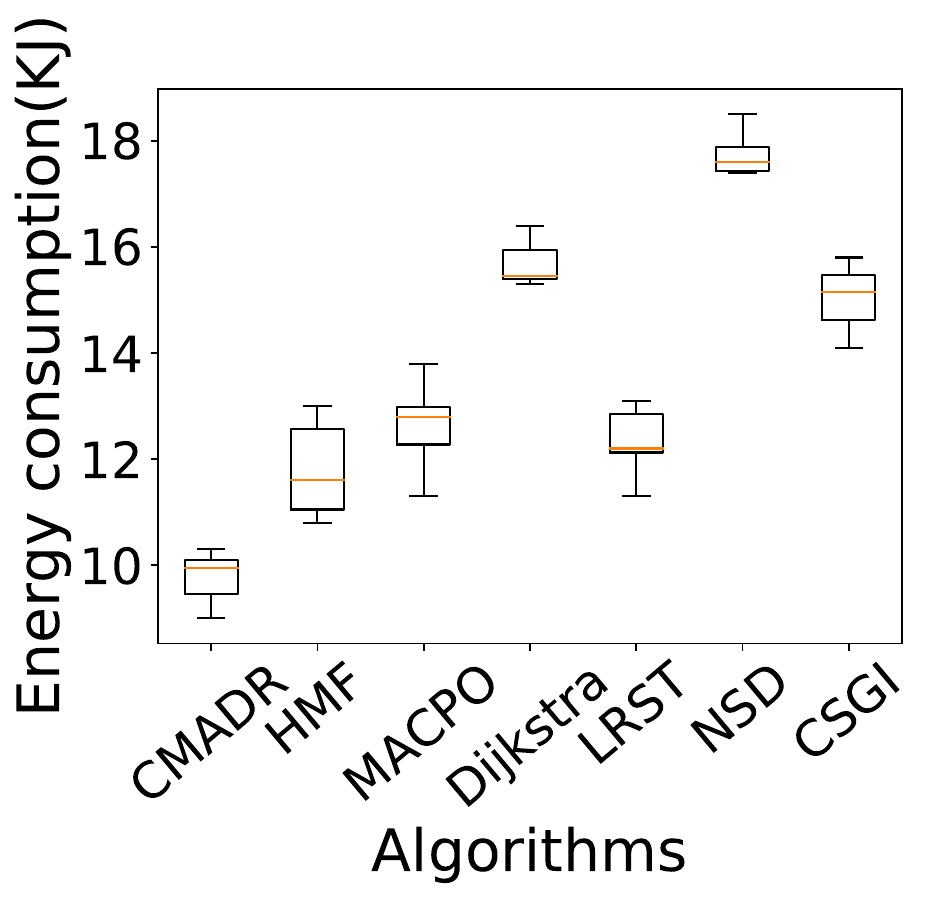}
\end{minipage}%
}%
\subfigure[Packet loss rate]{
\begin{minipage}[t]{0.33\linewidth}
\centering
\includegraphics[height=4cm,width=4.5cm]{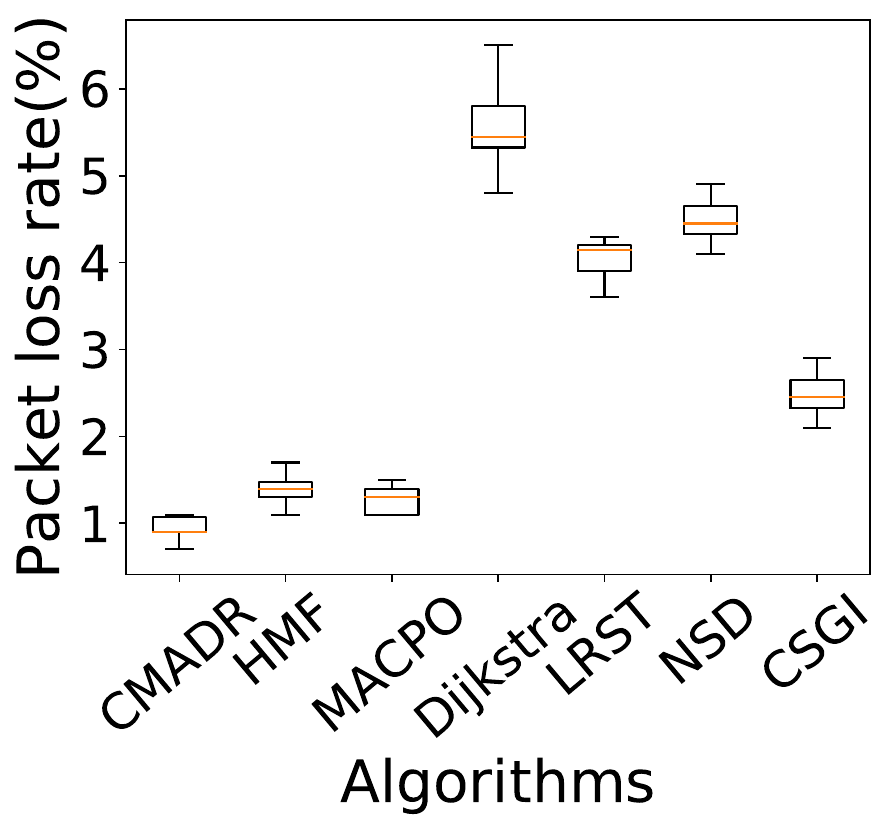}
\end{minipage}%
}%
\caption{(a) The average delay, (b) the energy consumption and (c) the packet loss rate for all algorithms were evaluated in 100 tests based on Oneweb.   CMADR algorithm showcases superior performance when compared to other methods.}
\vspace{-0.3cm}
\end{figure*}

\subsection{Ablation Study}
\begin{figure*}  
\vspace{-10pt}
\subfigure[Reward]{
\begin{minipage}[t]{0.25\linewidth}
\centering
\includegraphics[height=3.5cm,width=4cm]{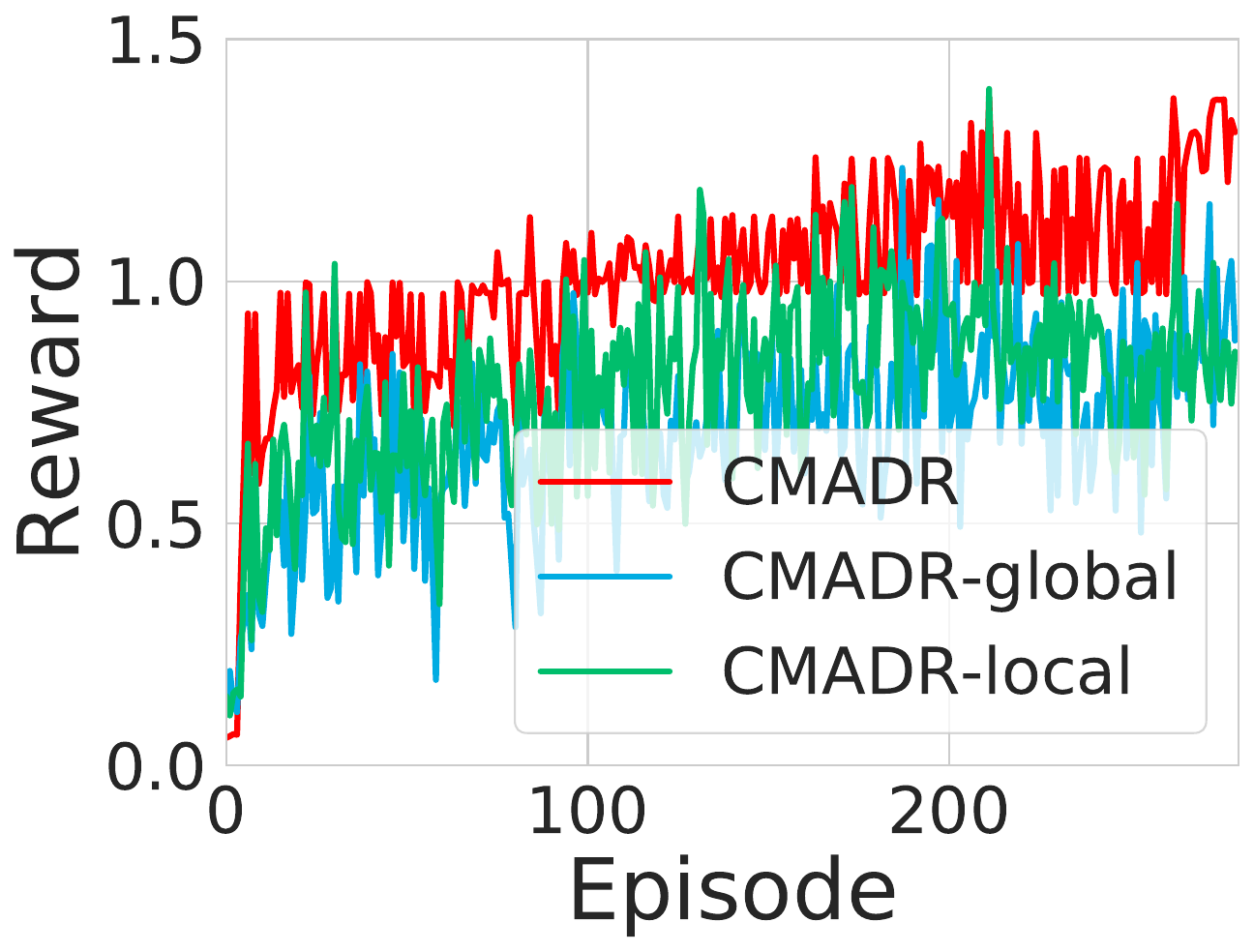}
\end{minipage}%
}%
\subfigure[Average delay]{
\begin{minipage}[t]{0.25\linewidth}
\centering
\includegraphics[height=3.5cm,width=4cm]{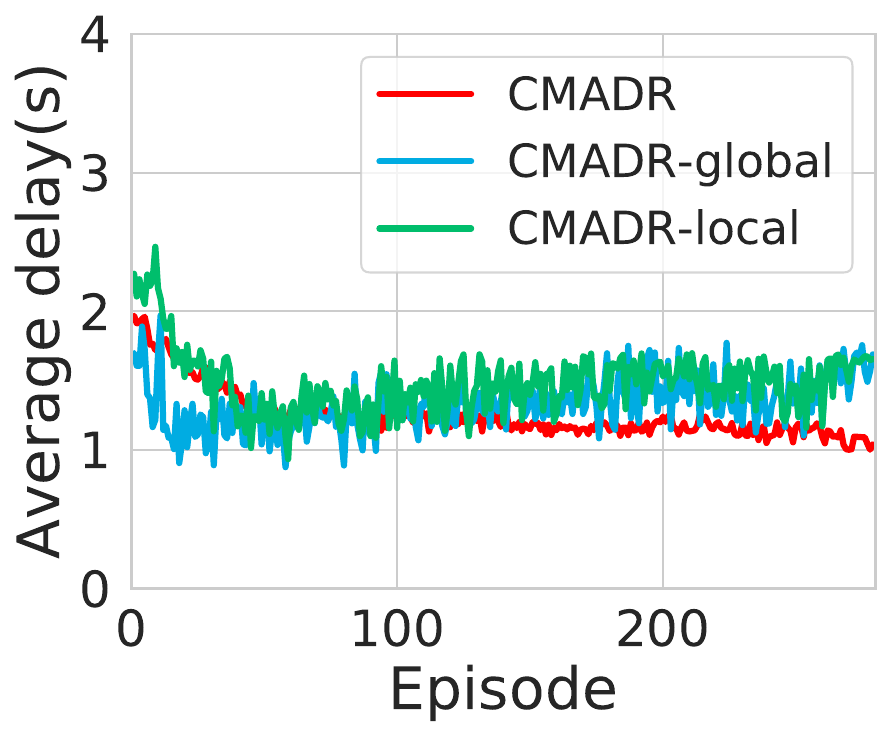}
\end{minipage}%
}%
\subfigure[Energy consumption]{
\begin{minipage}[t]{0.25\linewidth}
\centering
\includegraphics[height=3.5cm,width=4cm]{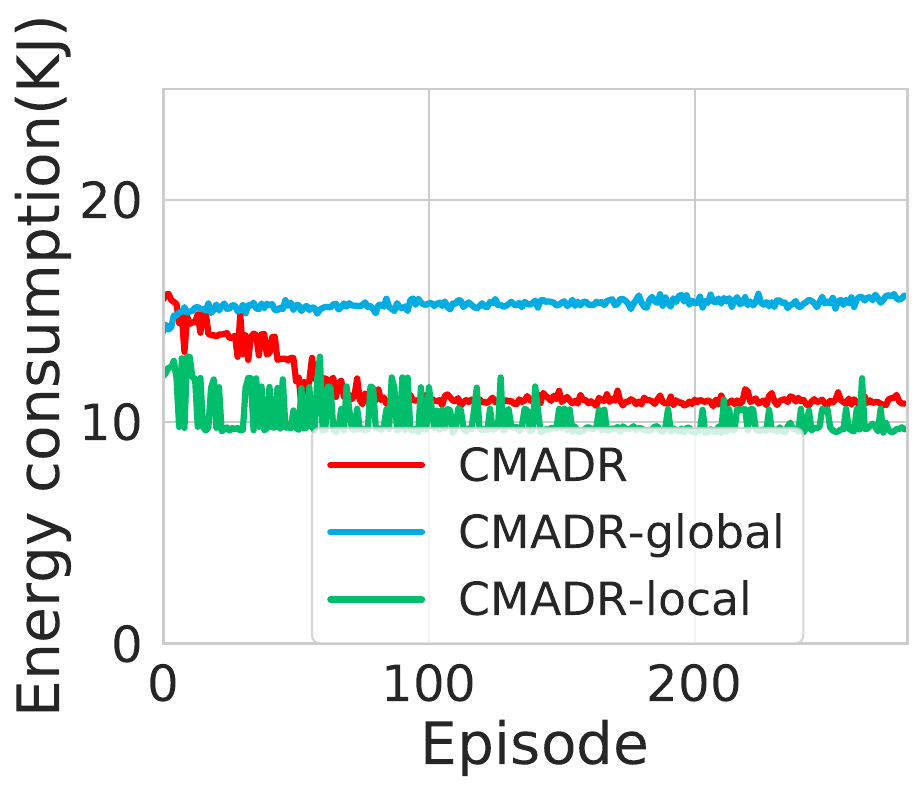}
\end{minipage}%
}%
\subfigure[Packet loss rate]{
\begin{minipage}[t]{0.25\linewidth}
\centering
\includegraphics[height=3.5cm,width=4cm]{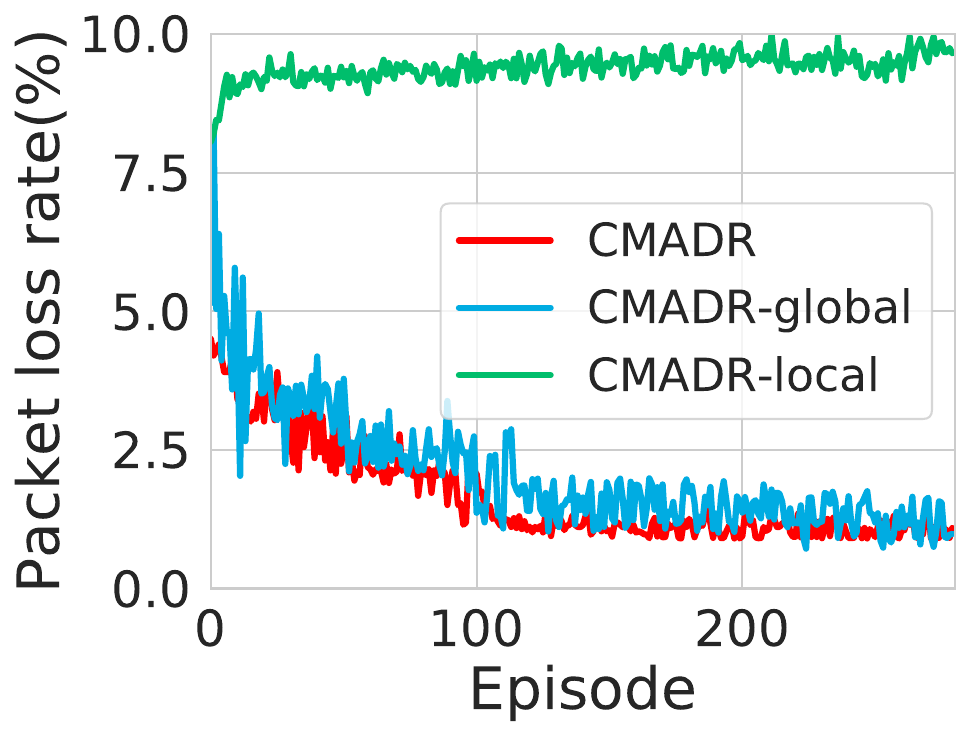}
\end{minipage}%
}%
\caption{For Telesat, CMADR, CMADR-global, and CMADR-local can ultimately converge while only CMADR effectively balances the average delay and energy consumption limits, as well as the packet loss rate limit, simultaneously.}
\vspace{-0.3cm}
\end{figure*}

\begin{figure*}  
\vspace{-3pt}
\subfigure[Reward]{
\begin{minipage}[t]{0.25\linewidth}
\centering
\includegraphics[height=3.5cm,width=4cm]{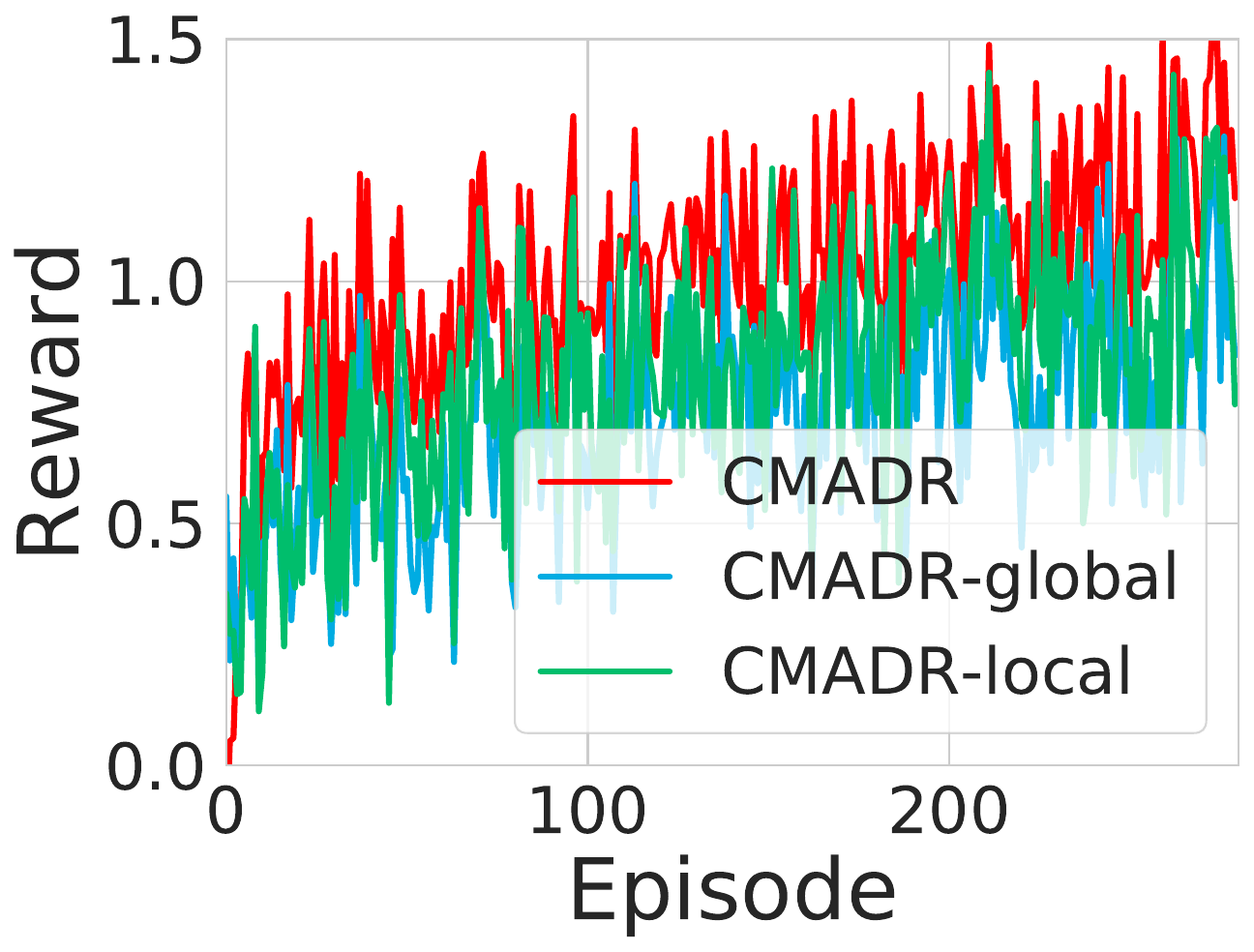}
\end{minipage}%
}%
\subfigure[Average delay]{
\begin{minipage}[t]{0.25\linewidth}
\centering
\includegraphics[height=3.5cm,width=4cm]{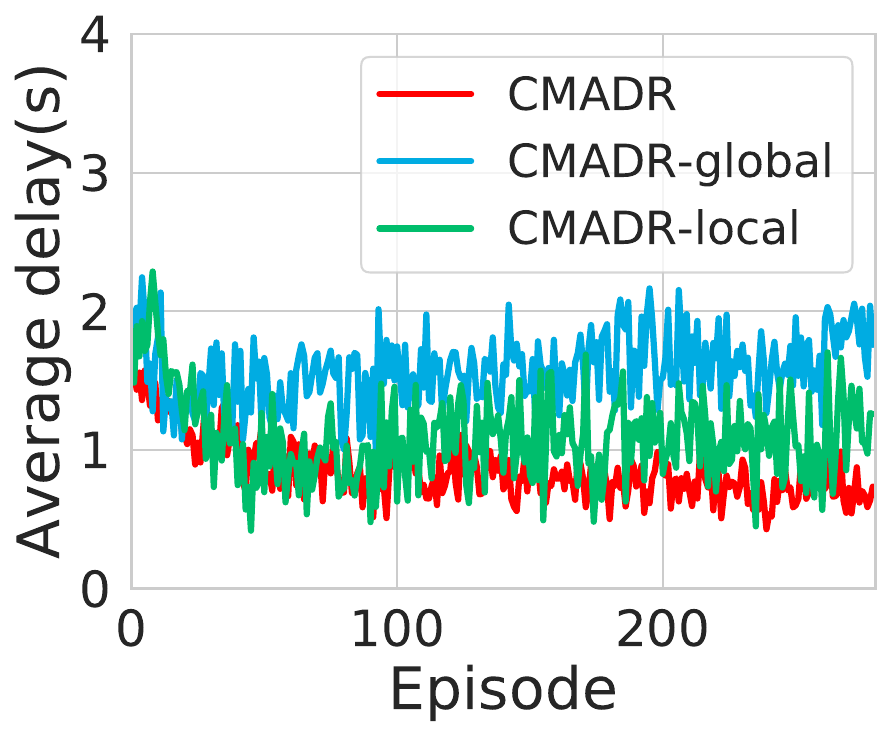}
\end{minipage}%
}%
\subfigure[Energy consumption]{
\begin{minipage}[t]{0.25\linewidth}
\centering
\includegraphics[height=3.5cm,width=4cm]{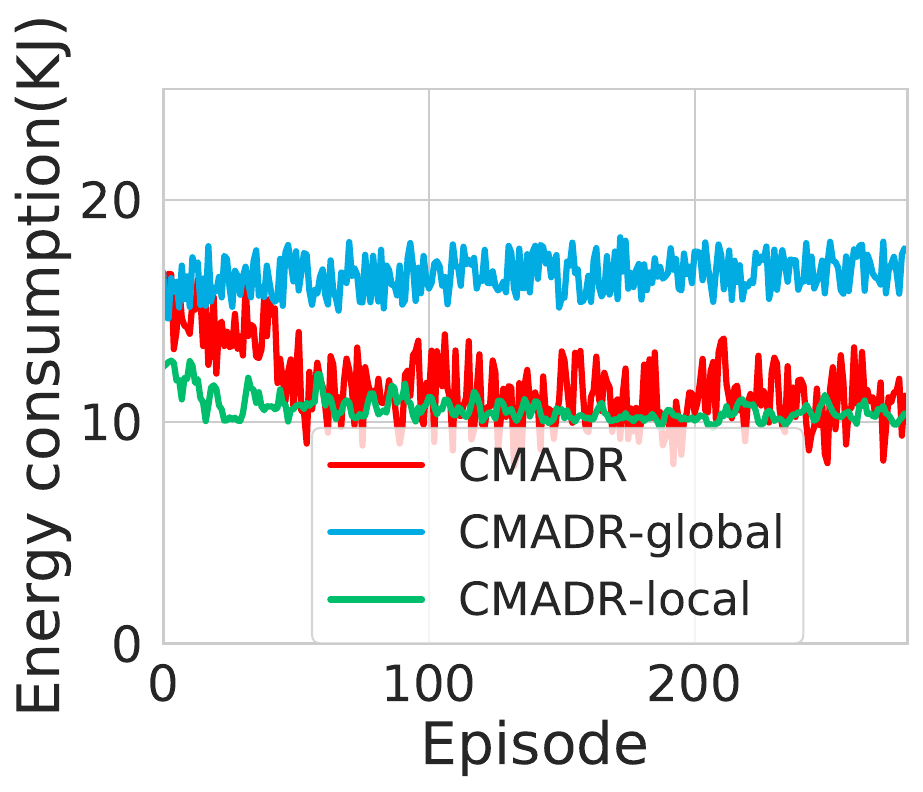}
\end{minipage}%
}%
\subfigure[Packet loss rate]{
\begin{minipage}[t]{0.25\linewidth}
\centering
\includegraphics[height=3.5cm,width=4cm]{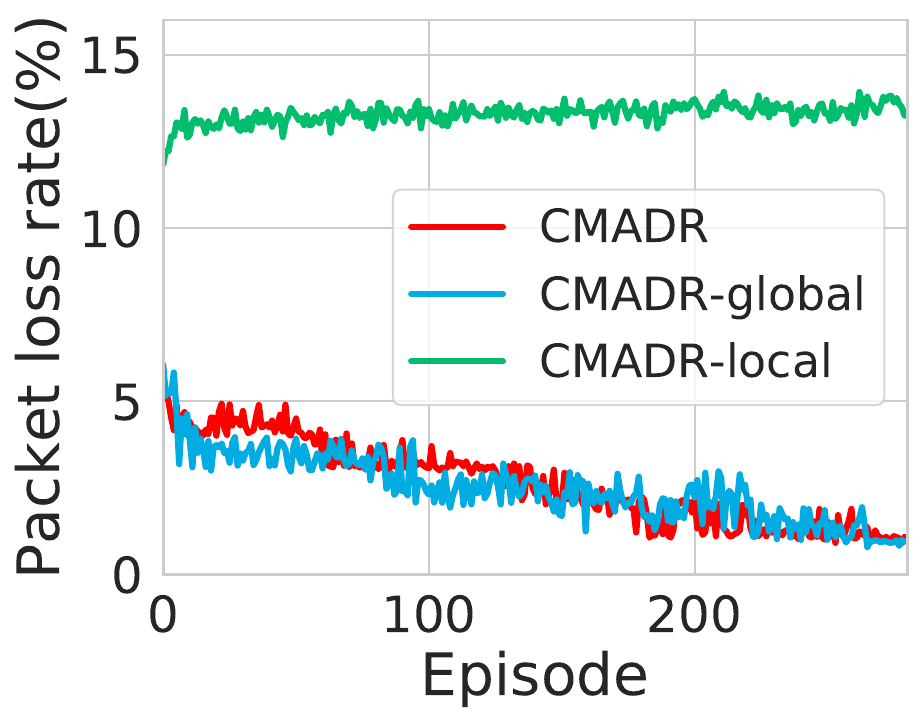}
\end{minipage}%
}%
\caption{For OneWeb, CMADR-global ensures the fulfillment of the packet loss rate constraint but falls short in meeting the energy consumption constraint, whereas CMADR-local behaves in the opposite manner. Both of them do not possess the same level of expertise as CMADR in achieving overall balance.}
\vspace{-0.3cm}
\end{figure*}

To illustrate the significance of the global cost critic and local cost critics in meeting corresponding constraints, we performed an ablation study that involved a comparison of three pertinent algorithms: \textbf{CMADR}, \textbf{CMADR-global}, and \textbf{CMADR-local}. \textbf{CMADR} represents the comprehensive routing algorithm we proposed, which incorporates both the global packet loss constraint and local energy consumption constraints. On the other hand, \textbf{CMADR-global} solely focuses on the global packet loss constraint, omitting the considerations of local energy consumption constraints. In contrast, \textbf{CMADR-local} exclusively addresses the local energy consumption constraints, without considering the global packet loss constraint.

The corresponding curves for the Telesat and OneWeb constellations are illustrated in Figure 8 and Figure 9 respectively. In Figure 8 (a) and Figure 9 (a), it can be observed that the reward curve of all three algorithms converges.  By examining Figure 8 (c), (d) and Figure 9 (c), (d), it becomes evident that CMADR-global can only guarantee the satisfaction of the packet loss rate constraint while failing to meet the constraint on energy consumption. On the other hand, CMADR-local exhibits the opposite behavior, satisfying the constraint on energy consumption but falling short in meeting the packet loss rate constraint. Interestingly, both the CMADR-global and CMADR-local algorithms in Figure 8 (b) and Figure 9 (b) show a slightly faster decrease in the average delay compared to CMADR alone. However, this temporary improvement does not persist in later stages. The reason behind this could be that unrestricted energy consumption and packet loss rates actually increase the probability of network congestion, leading to a less favorable overall average packet delay. 

The characteristics of all the comparative methods mentioned earlier, as well as those of the ablation experiment's comparative methods, are summarized in Table III. The Dijkstra, LRST, NSD, and CSGI algorithms are static, as they can pre-plan the routes on the ground without considering the dynamic link conditions. In contrast, our CMADR, CMADR-global, and CMADR-local algorithms dynamically plan routes based on the link environment's real-time status. CMADR-global focuses solely on the packet loss rate constraint, while CMADR-local considers only the global energy consumption constraint. CMADR, on the other hand, takes both constraints into account.
%\vspace{-10pt}
\begin{table}[htbp]
	\centering
	 
	\caption{SUMMARY OF  ALGORITHMS.}
  \begin{tabular}{|c|c|c|c|}
   \hline
   & & & \\[-6pt]
    Routing scheme     &  \makecell[c]{\textbf{S}tatic/\textbf{D}ynamic    }&
    \makecell[c]{GLobal critic   }
    &\makecell[c]{Local critic } \\
     & & & \\[-6pt]
    \hline
     & & & \\[-6pt]

    Dijkstra~\cite{553679}    &  S & -   &  -  \\    

    LRST~\cite{7275422} & S&  -  &   -  \\

    NSD~\cite{NSD} &  S &  -& -  \\
    
    CSGI~\cite{9796886}    &  S & - &  -   \\   

    %\textcolor{blue}{HMF~\cite{10012942}}    &  \textcolor{blue}{D} & \textcolor{blue}{\Checkmark} &  \textcolor{blue}{\Checkmark}   \\  
    %\textcolor{blue}{MACPO~\cite{GU2023103905}}    &  \textcolor{blue}{D} & \textcolor{blue}{\Checkmark} &  \textcolor{blue}{\Checkmark}   \\   
    HMF~\cite{10012942}    &  D & \Checkmark & \Checkmark   \\  
    MACPO~\cite{GU2023103905}    &  D & \Checkmark &  \Checkmark   \\   
    \hline
    CMADR  & D & \Checkmark  & \Checkmark    \\

    CMADR-global  & D & \Checkmark  & \XSolid    \\  
     
    CMADR-local   & D & \XSolid  & \Checkmark    \\  
    
    \hline

  \end{tabular}

  \label{sample-table}
\end{table}
%\vspace{-10pt}

\subsection{Algorithm  Parameters Analysis}

\begin{figure*}  

\subfigure[Reward]{
\begin{minipage}[t]{0.25\linewidth}
\centering
\includegraphics[height=3.5cm,width=4cm]{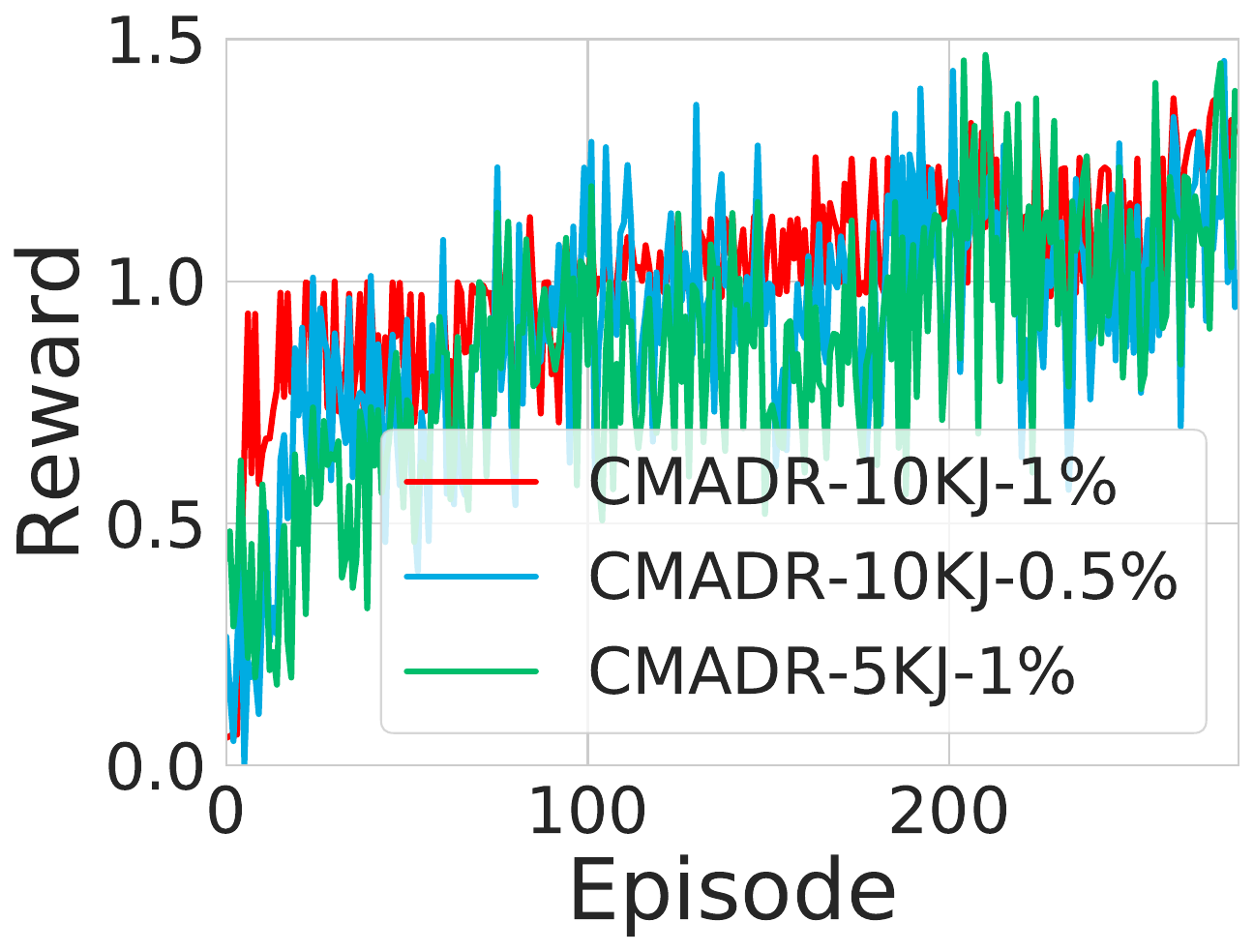}
\end{minipage}%
}%
\subfigure[Average delay]{
\begin{minipage}[t]{0.25\linewidth}
\centering
\includegraphics[height=3.5cm,width=4cm]{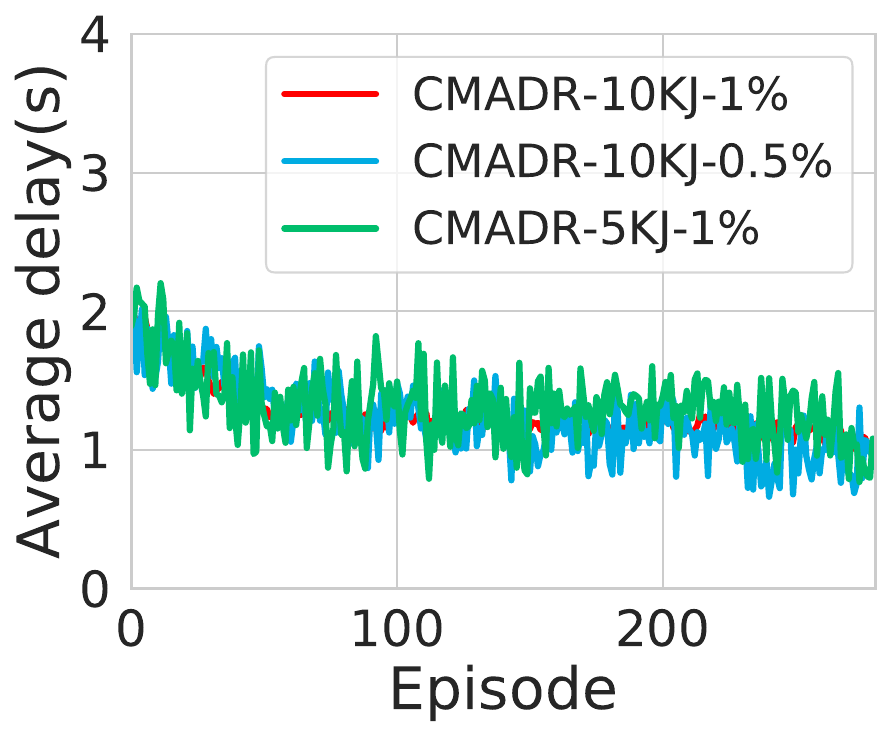}
\end{minipage}%
}%
\subfigure[Energy consumption]{
\begin{minipage}[t]{0.25\linewidth}
\centering
\includegraphics[height=3.5cm,width=4cm]{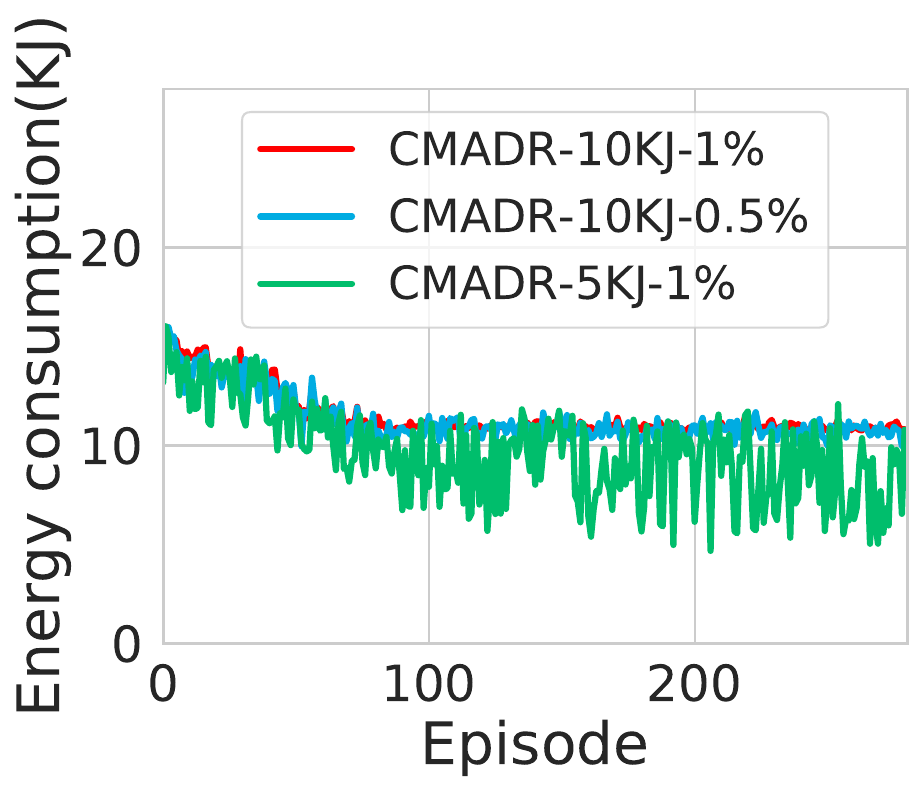}
\end{minipage}%
}%
\subfigure[Packet loss rate]{
\begin{minipage}[t]{0.25\linewidth}
\centering
\includegraphics[height=3.5cm,width=4cm]{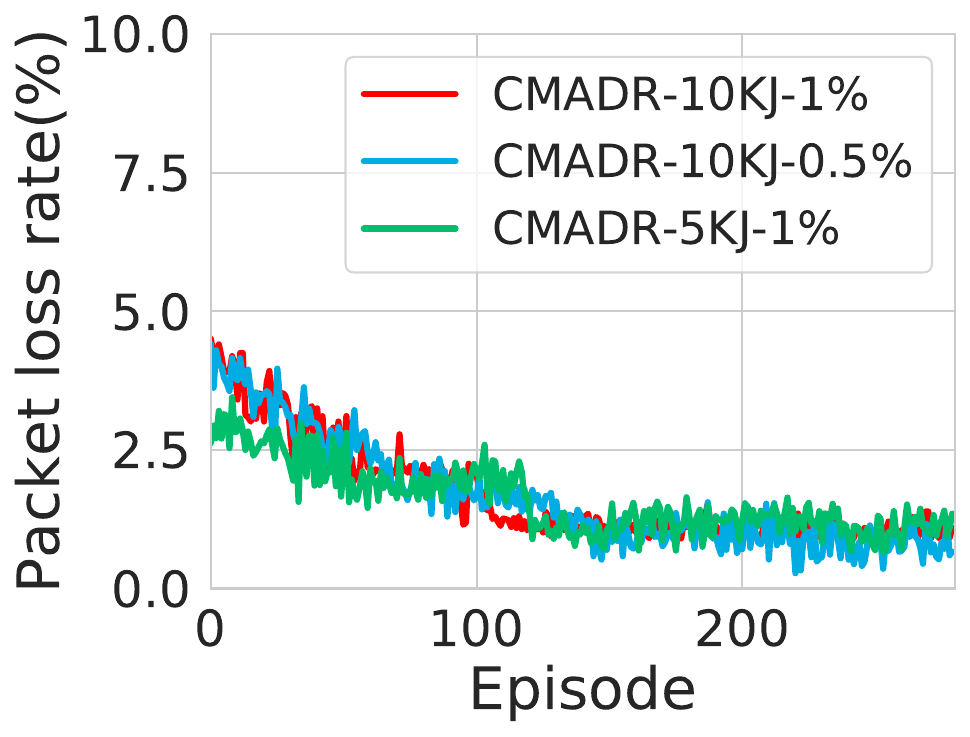}
\end{minipage}%
}%
\caption{For Telesat, by setting different thresholds for energy consumption and packet loss rate constraints individually, CMADR can effectively optimize the average delay while satisfying these different constraints.}
\vspace{-0.3cm}
\end{figure*}

\begin{figure*}  
\vspace{-3pt}
\subfigure[Reward]{
\begin{minipage}[t]{0.25\linewidth}
\centering
\includegraphics[height=3.5cm,width=4cm]{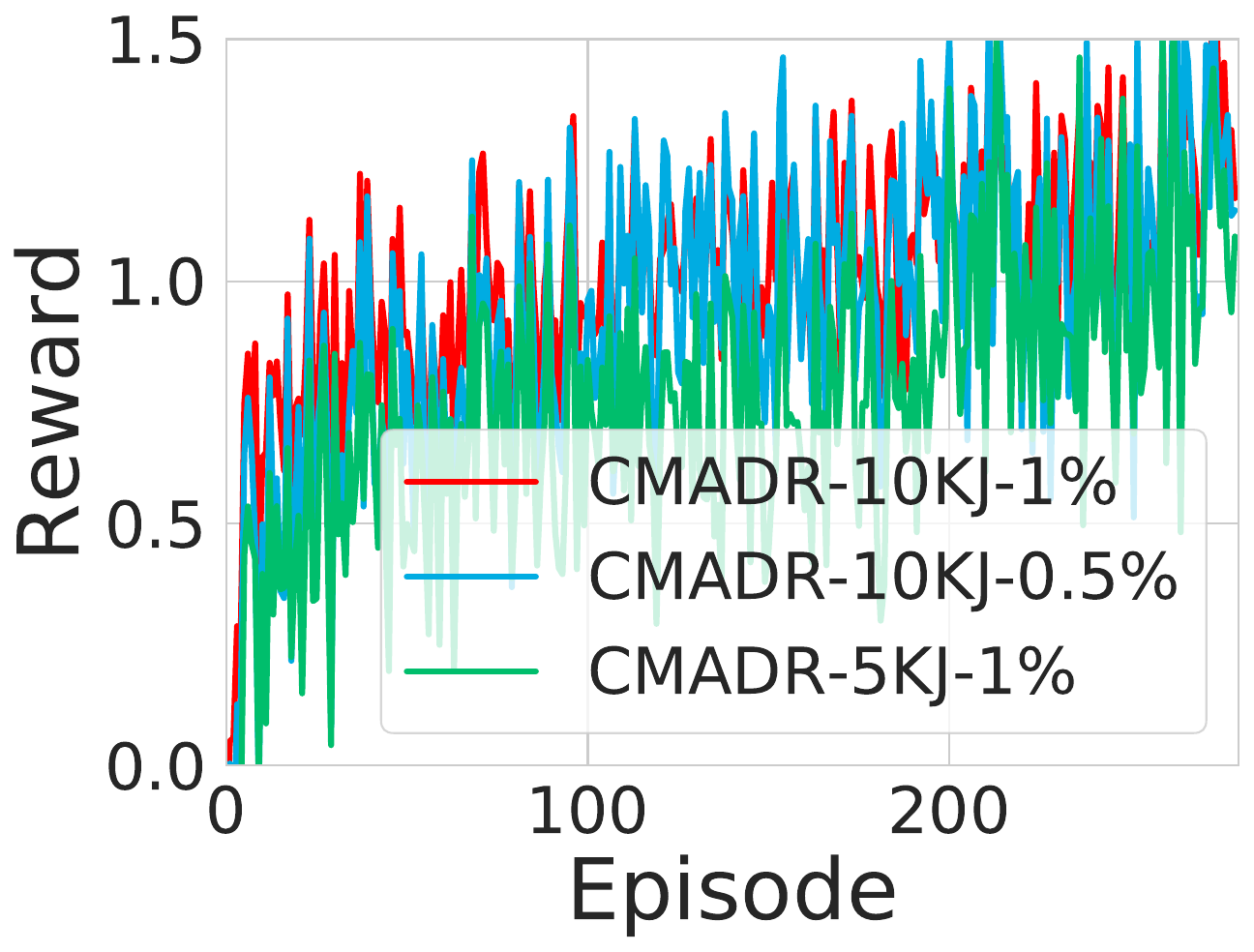}
\end{minipage}%
}%
\subfigure[Average delay]{
\begin{minipage}[t]{0.25\linewidth}
\centering
\includegraphics[height=3.5cm,width=4cm]{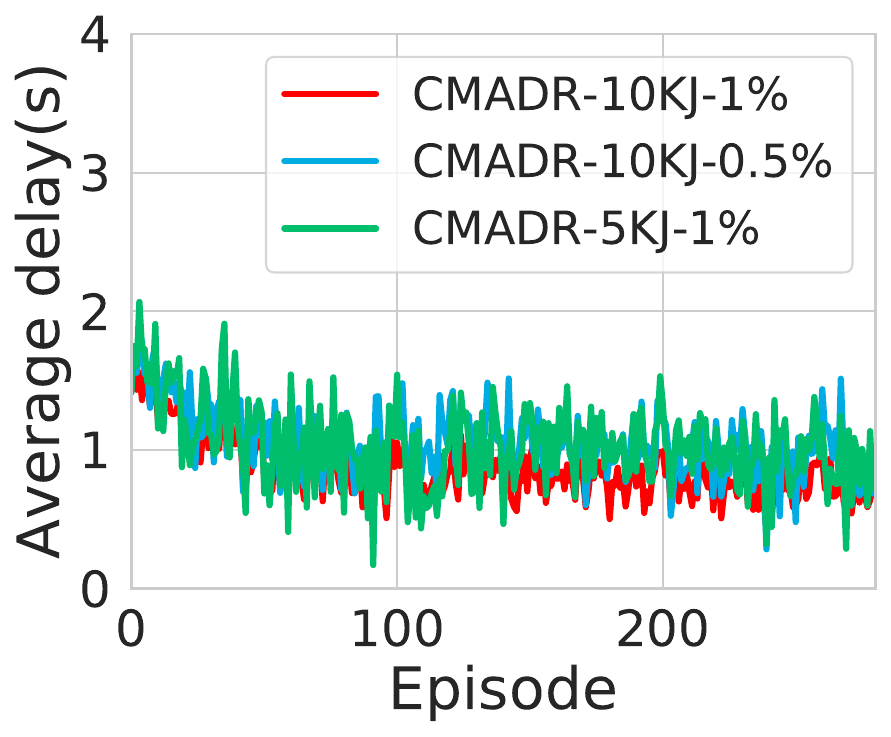}
\end{minipage}%
}%
\subfigure[Energy consumption]{
\begin{minipage}[t]{0.25\linewidth}
\centering
\includegraphics[height=3.5cm,width=4cm]{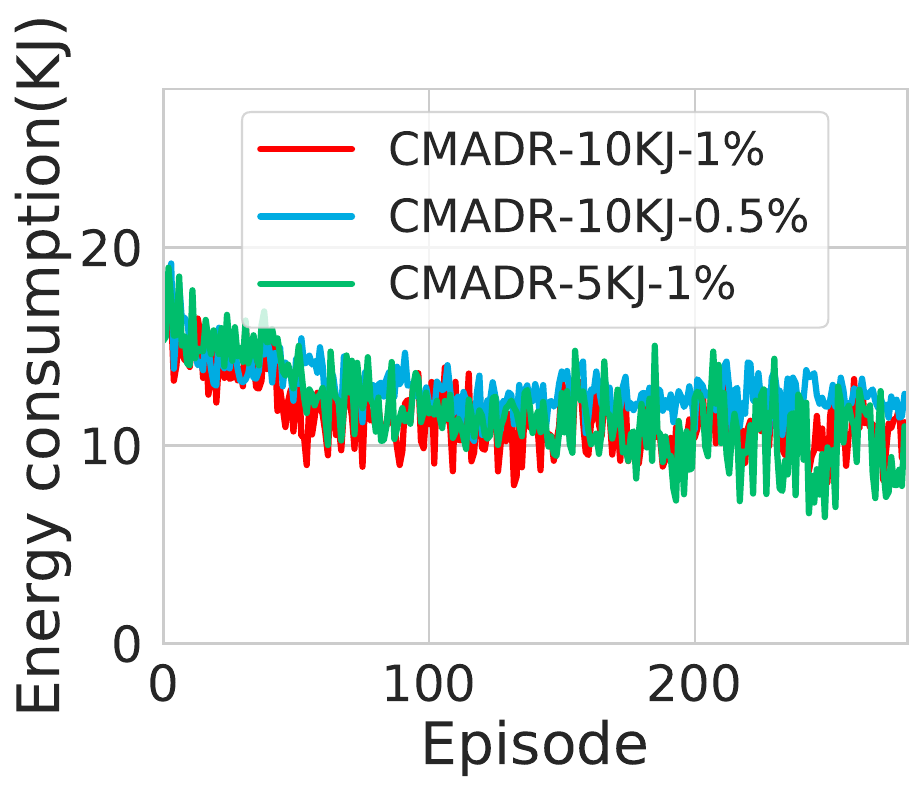}
\end{minipage}%
}%
\subfigure[Packet loss rate]{
\begin{minipage}[t]{0.25\linewidth}
\centering
\includegraphics[height=3.5cm,width=4cm]{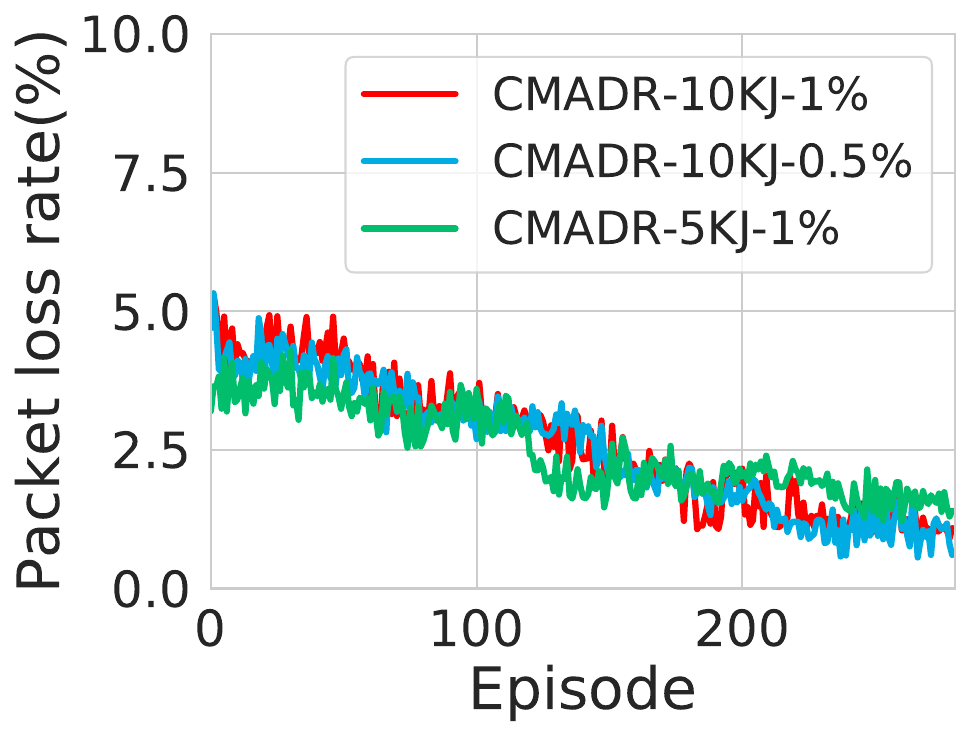}
\end{minipage}%
}%
\caption{In the context of OneWeb, CMADR also demonstrates its capability to optimize the average delay while effectively meeting the constraints even under different configurations.}
\vspace{-0.3cm}
\end{figure*}

To demonstrate the impact of different constraint thresholds on algorithm performance, we initially maintain a constant threshold for energy consumption and vary the threshold values for packet loss rate, setting it to 0.5\%. The corresponding result curves are presented in Figure 10 and Figure 11 for the Telesat and OneWeb constellations, respectively. In Figure 10 (d) and Figure 11 (d), we observe that the curves continue to show a decreasing trend. However, as the training progresses, it becomes difficult for them to further decrease and reach the relatively stringent requirement of 0.5\% packet loss rate. Additionally, corresponding energy consumption values are higher, and the average delay is also larger. This indicates that the system faces challenges in optimizing the objectives while satisfying all the constraints simultaneously.

Similar conclusions can be drawn from Figure 10 and Figure 11, where we maintain a constant threshold for packet loss rate and vary the threshold values for energy consumption, setting it to 5KJ. Once again, the tighter energy consumption constraint makes it difficult for them to effectively optimize average packet delay while ensuring a low packet loss rate.

\section{CONCLUSION}

In this study, we propose an ISTN system that integrates satellites and ground stations for collaborative packet routing, aiming to minimize average packet delay while ensuring energy efficiency and meeting packet loss rate constraints. We formulate this as a  max-min problem and introduce the CMADR routing algorithm, which ensures that the strategy updates can meet the constraints while optimizing the objective, as confirmed by extensive simulations.

\bibliographystyle{IEEEtran}
\bibliography{ref}
\end{document}